\documentclass[aps,nofootinbib,notitlepage,longbibliography,twocolumn, superscriptaddress]{revtex4-1}
\usepackage{amsmath,amssymb}
\usepackage{slashed}
\usepackage{graphicx}
\usepackage{bm}
\usepackage{float}
\usepackage[T1]{fontenc}
\usepackage[utf8]{inputenc}
\usepackage{gauss} 
\usepackage[top=1.0in,bottom=1.0in,left=1.0in,right=1.0in]{geometry}
\usepackage[colorlinks,linkcolor=blue,citecolor=blue]{hyperref}
\usepackage{subfig}
\usepackage{multirow}
\newcommand{\be}{\begin{equation}}
\newcommand{\ee}{\end{equation}}
\newcommand{\bea}{\begin{eqnarray}}
\newcommand{\eea}{\end{eqnarray}}
\newcommand{\ba}{\begin{eqnarray}}
\newcommand{\ea}{\end{eqnarray}}

\allowdisplaybreaks

\def\be{\begin{eqnarray}}
\def\ee{\end{eqnarray}}
\def\bea{\be}
\def\eea{\ee}

\def\roughly#1{\mathrel{\raise.3ex\hbox{$#1$\kern-.75em%
\lower1ex\hbox{$\sim$}}}}

\usepackage{lipsum}

\date{\today}
\begin{abstract}
The gravitational form factors of light nuclei are evaluated up to momenta of the order of the nucleon mass, using the impulse approximation. The nucleon gravitational form factors are reduced non-relativistically, 
and used to derive the gravitational form factors of light nuclei. The deuteron gravitational form factors are analysed using the Reid soft core potential. The helium-4 gravitational form factors are assessed using the K-harmonics method, and  compared to those following from a  mean-field approximation with a Woods-Saxon potential. The importance of removing the center of mass motion for the ensuing form factors is emphasized.
The mass radii of these light nuclei are extracted and compared to their charge radii counterparts. The details of their pressure and shear distributions  
are discussed.
\end{abstract}
\begin{document}
\title{Gravitational form factors of light nuclei:\\ Impulse approximation
}
\author{Fangcheng He}
\email{fangcheng.he@stonybrook.edu}
\affiliation{Center for Nuclear Theory, Department of Physics and Astronomy,
Stony Brook University, Stony Brook, New York 11794–3800, USA}

\author{Ismail Zahed}
\email{ismail.zahed@stonybrook.edu}
\affiliation{Center for Nuclear Theory, Department of Physics and Astronomy,
Stony Brook University, Stony Brook, New York 11794–3800, USA}
\maketitle

\section{Introduction}
The gravitational form factors of the nucleon carry important information on its mass distribution, most of which is carried by constituent gluons. Recently, threshold photo-production of charmonium 
at JLab~\cite{GlueX:2019mkq}, has opened the possibility of measuring the gluonic component of the nucleon gravitational form factors. 
The high statistics results reported by the E12-007-collaboration~\cite{Duran:2022xag}, suggest
smaller mass radii for the proton in comparison to its electromagnetic radius.

Threshold electromagnetic production of charmonium off light nuclei, could open the possibility of understanding 
the nuclear effects on the gravitational form factors. The nucleus
is a collection of nucleons (protons and neutrons) bound by strong QCD interactions. Most of what is known about nuclei has been gleaned using electromagnetic probes at intermediate energies~\cite{Donnelly:1975ze}, where the nucleons appear as rigid but extended bodies exchanging mesons, albeit mostly pions~\cite{Riska:1989bh} (and references therein). The disparity between the fundamental and unconfined degrees of QCD (quarks and gluons) and the observed but confined 
degrees of freedom (mesons and nucleons) call for novel probes. The ultimate goal is to understand the composition of the nucleons, and how the nuclear interactions emerge in a nucleus.

The difficult character of the strong nuclear interaction, has required the use of approximate models to account for the motion of the nucleons in a bound nucleus. Mean field models of which the shell model is the ultimate realization, have proven successful in interpreting many aspects of low and intermediate nuclear physics. However, much is  still needed for a theory to be sufficiently accurate and predictive. For this reason, the study of simpler nuclei such as deuterium, triton and helium-3,4 should prove useful for the study of novel probes, such as the one provided by the energy-momentum tensor (EMT).

The simplest nuclear system is of course the deuteron. Its binding energy
(2.225 MeV), 
charge radius (2.13 fm)  and magnetic moment (0.857 in Bohr magnetons) are well established, which strongly constrain the pair nucleon-nucleon interaction~\cite{Brown1974NucleonnucleonI}. The deuteron large size and weak binding, suggests that the nuclear interaction is due to single pion exchange between almost on-shell nucleons. The deuteron is a diffuse nucleus.

The purpose of this work is to provide the starting framework for the nuclear effect on the deuteron EMT. 
We will derive in detail its gravitational form factors using the impulse approximation. The results are readily extended to spherically symmetric and light nuclei
such as helium-4, which is the prototype nucleus per excellence, given  that its binding energy per particle is close to the saturation one.

For completeness, we note that 
 the gravitational D-form factors for nuclei were initially discussed   using a liquid drop model in~\cite{Polyakov:2002yz}, relativistic nuclear potentials in~\cite{Guzey:2005ba}, nuclear structure~\cite{Kim:2012ts,Hudson:2017xug,Dupre:2015jha,Dong:2013wka,Scopetta:2004kj,Scopetta:2004ex}
and more recently  the generalized Skyrme model in~\cite{GarciaMartin-Caro:2023klo}.  
Also, the spatial densities and forces in the deuteron, were discussed in~\cite{Freese:2022yur} using the
light cone convolution model. An estimate of the mass radius of
 helium-4,  was suggested recently using $\phi$-meson photoproduction~\cite{Wang:2023uek}. 

The outline of this paper is as follows: in section~\ref{SEC2} we briefly review the chief aspects of the deuteron S,D contributions using Reid soft core potential.
In section~\ref{SEC3} we summarize the relevant aspects of the nucleon gravitational A,B,C=$\frac 14$ D form factors. 
To use them for low and intermediate energies up to the nucleon mass scale, we explicitly present their non-relativistic expansions. 
In section~\ref{SEC4} we derive the deuteron 
gravitational form factors in the impulse approximation, and in leading order in the
recoil momentum of the spectator nucleon. 
In section~\ref{SEC5} we extend our results to helium-4, using both the K-harmonics method, and the mean-field approximation with a Woods-Saxon potential. The importance of removing the spurious center of mass motion,
while addressing the form factors of light nuclei
is emphasized. In section~\ref{SEC6} we detail the extraction of the mass radii from the pertinent EMT form factors. Our conclusions are in section~\ref{SEC7}. In Appendix~\ref{APPA}, we 
compare our deuteron and helium-4 charge form factors, versus the existing
data. In Appendix~\ref{app:pre}, we briefly recall the pressure and shear force used.

\section{Deuteron state}
\label{SEC2}
The deuteron with the tiny 2.2 MeV binding, is a loosely bound light nucleus, composed of almost quasi-free
proton plus neutron held together by a long range pion-exchange interaction. In the non-relativistic approximation,
the deuteron wavefunction is a mixture of ${}^3S_1+{}^3D_1$,
\be
\label{D1X}
\Phi_{m}(r)=
\bigg(\frac ur+\frac 1{\sqrt 8}\frac wr\,S_{12}\bigg)\,\frac{\chi_m}{\sqrt{4\pi}}
\ee
with the deuteron quadrupole operator
\bea
S_{12}=6S\cdot \hat r S\cdot \hat r-2 S^2=6Q^{ij}\hat r^i\hat{r}^j
\eea
with total spin $\vec S$, where $Q^{ij}$ is the quadrupole operator $Q^{ij}=\frac{1}{2}(S^iS^j+S^jS^i)-\frac{2}{3}\delta^{ij}$.
The reduced radial wavefunctions $u,w$  are normalized,
\be
 \int_0^\infty dr(u^2+w^2)=1
\ee

\begin{figure}
\centering
    \includegraphics[height=5cm,width=0.9\linewidth]{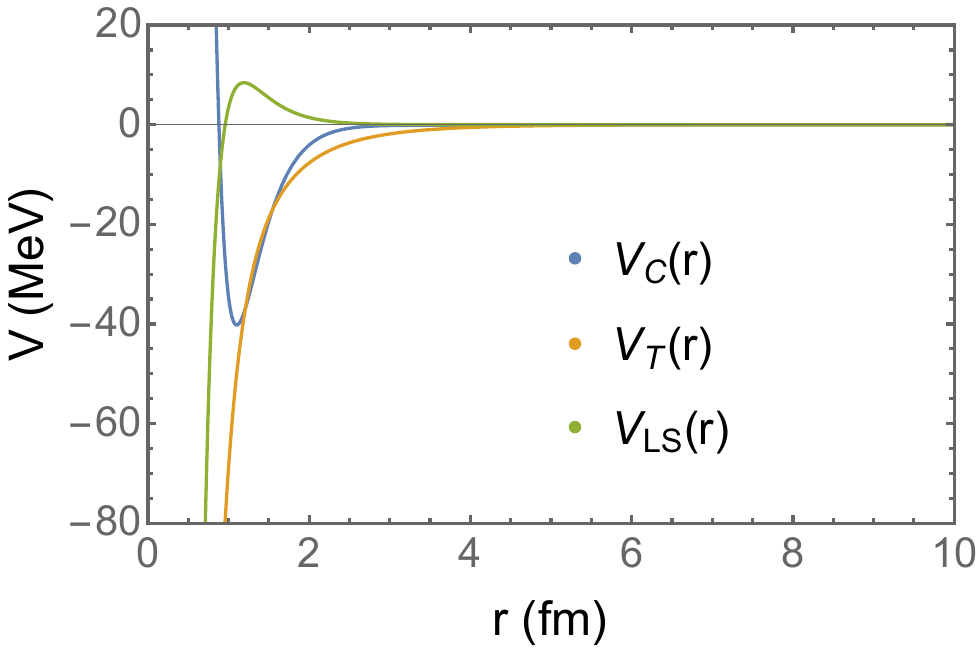}
    \includegraphics[height=5cm,width=0.9\linewidth]{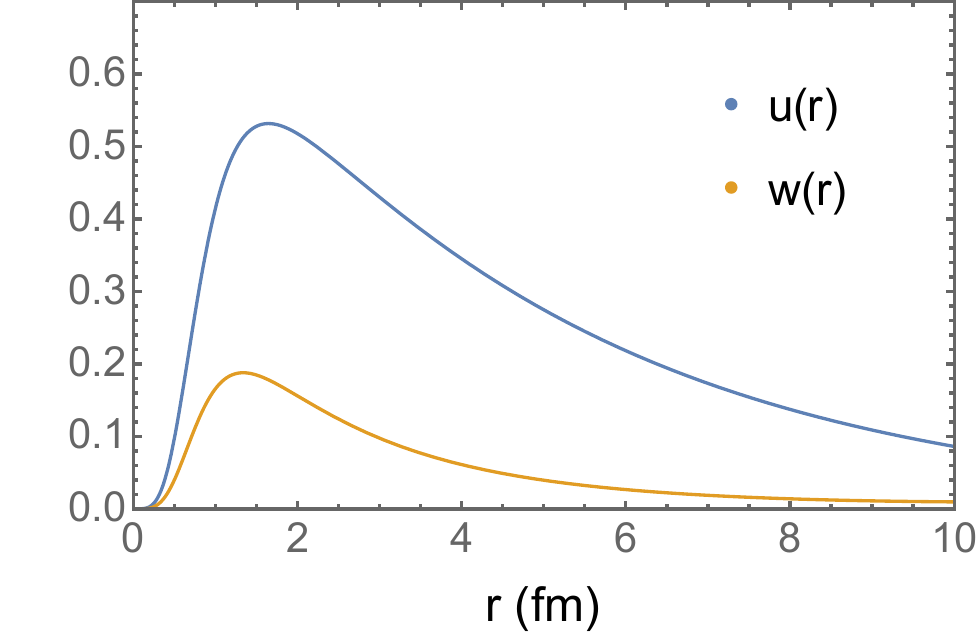}
    \caption{Top: Reid soft core potentials $V_{C,T,LS}$;
    Bottom: Deutron S,D wavefunctions in Eq.~(\ref{eq:sch}).}
    \label{fig:REID}
\end{figure}

The coupled $u,w$ reduced radial components 
${}^3S_1$ and ${}^3D_1$ of the deuteron wavefunction, will be sought using a central and tensor interaction~\cite{Reid:1968sq}
\begin{widetext}
\bea\label{eq:sch}
&&u{''}+\frac m{\hbar^2}(-E-V_C(r))\,u-\sqrt{8}\frac m{\hbar^2}V_T\,w=0,\nonumber\\
&&w^{''}+\frac m{\hbar^2}\bigg(-E-V_C(r)-\frac {6\hbar^2}{mr^2}+2V_T(r)+3V_{LS}(r)\bigg)\,w
-\sqrt{8}\frac m{\hbar^2}V_T\,u=0
\eea
with the Reid soft core potential
$V_R=V_C+V_TS_{12}+V_{LS}\,L\cdot S$,
\be\label{eq:potential}
\label{REID1}
V_C&=&-h\frac{e^{-x}}{x}+105.468\,({\rm MeV})\frac{e^{-2x}}x-3187.8\,({\rm MeV})\frac{e^{-4x}}x+9924.3\,({\rm MeV})\frac{e^{-6x}}x,
\nonumber\\
V_T&=&-h\bigg(\bigg(\frac 1x +\frac 3{x^2}+\frac 3{x^3}\bigg)\,e^{-x}-\bigg(\frac {12}{x^2}+\frac 3{x^3}\bigg)\,e^{-4x}\bigg)+351.77 \,({\rm MeV})\frac{e^{-4x}}x-1673.5\,({\rm MeV})\frac{e^{-6x}}{x},\nonumber\\
V_{LS}&=&708.91\,({\rm MeV})\frac{e^{-4x}}{x}-2713.1\,({\rm MeV})\frac{e^{-6x}}x.
\ee
\end{widetext}
Here $x=\mu r$ is the pion range fixed by $\mu=0.7\,\text{fm}^{-1}$, as illustrated in Fig.~\ref{fig:REID} (top). Also 
$h=10.463\,{\rm MeV}$ and $\hbar^2/m$ is assumed to be 41.47 MeV\,fm$^2$ with $m$ the twice reduced mass of proton and neutron. The numerical S- and D-wavefunctions solution to the coupled equations (\ref{REID1}) and valid for $x<10.01$,  are shown in Fig~\ref{fig:REID} (bottom). 
For $x>10.01$, the explicit solutions are~\cite{Reid:1968sq},
\bea
u(r)&=&0.87758e^{-\alpha \mu r},\nonumber\\
w(r)&=&0.0023e^{-\alpha \mu r}\left(1+\frac{3}{\alpha \mu r}+\frac{3}{(\alpha \mu r)^2}\right),
\eea
with $\alpha=(mE_D)^{1/2}/(\mu\hbar)$. The deuteron solution in Fig~\ref{fig:REID} (bottom) carries  binding energy $E_D = 2.2246\,{\rm MeV}$, and a quadrupole moment
\bea
Q^E_D=\frac 14\int d^3r |\Phi_1(r)|^2\,(3z^2-r^2)\approx 0.31\,{\rm fm^2}
\eea
in the z-direction and  the maximally stretches spin state.
The deuteron is mostly cigar-shaped. This deformation amounts to $p_D=6.53\%$, the percentage of admixture of D-state in the deuteron~\cite{Reid:1968sq}.

\section{Nucleon EMT}
\label{SEC3}
The standard decomposition of EMT form factor in a nucleon state is~\cite{Pagels:1966zza,Carruthers:1971uy,Polyakov:2018zvc}
\begin{widetext}
\be
\label{A1tmp}
T_N^{\mu\nu}(p_2,p_1)=\left<p_2|T^{\mu\nu}(0)|p_1\right>=\overline{u}(p_2)\left(
A(k)\gamma^{(\mu}\bar{P}^{\nu)}+B(k)\frac{i\bar{P}^{(\mu}\sigma^{\nu)\alpha}k_\alpha}{2m_N}+C(k)\frac{k^\mu k^\nu-\eta^{\mu\nu}k^2}{m_N}\right)u(p_1)\,,
\ee
\end{widetext}
with  $a^{(\mu}b^{\nu)}=  \frac 12 (a^\mu b^\nu+a^\nu b^\mu)$, 
 $k^2=(p_2-p_1)^2=t$, $\bar{P}=(p_1+p_2)/2$  and the normalization $\overline u u=1$. 
(\ref{A1tmp}) is conserved and tracefull. Note that in other conventions, the C-form factor is also referred to by  $D(k)=4C(k)$. 

\begin{figure}
    \includegraphics[height=5cm,width=0.9\linewidth]{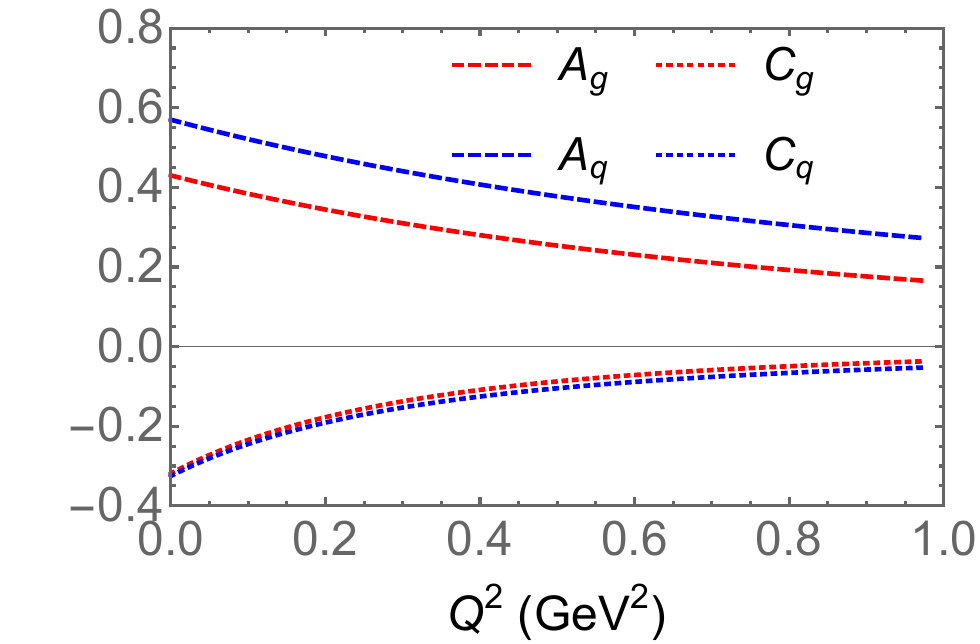}
    \caption{Nucleon GFFs: $A_q, C_q$ (quarks) are  from the recent lattice results~\cite{Hackett:2023rif}, and $A_g, C_g$ (gluons) are from the holographic model ~\cite{Mamo:2022eui}.}
    \label{fig:pressureQ}
\end{figure}

The gluonic gravitational form factors (GFFs) have been analyzed both analytically and numerically,
and more recently extracted empirically 
with overall good agreements. For the numerical analyses below, we will 
use a tripole parametrization of  the holographic results for the gluon GFFs,  and a dipole parametrization for the lattice quark GFFs with $k^2=-Q^2$ space-like,
\bea
\label{ABFF}
A_g(k)=\frac{A_g(0)}{\bigg(1+\frac {Q^2}{m^2_{TT,g}}\bigg)^3},\nonumber\\
A_q(k)=\frac{A_q(0)}{\bigg(1+\frac {Q^2}{m^2_{TT,q}}\bigg)^2},\nonumber\\
C_g(k)=\frac{\frac 14 D_g(0)}{\bigg(1+\frac {Q^2}{m^2_{SS,g}}\bigg)^3},\nonumber\\
C_q(k)=\frac{\frac 14 D_q(0)}{\bigg(1+\frac {Q^2}{m^2_{SS,q}}\bigg)^2}\nonumber\\
\eea
with $m_{TT,g}=1.612\,{\rm GeV}$, $m_{TT,q}=1.477(44)\,{\rm GeV}$, $m_{SS,g}=0.963\,{\rm GeV}$, $m_{SS,q}=0.81(14) \,{\rm GeV}$, $A_g(0)=0.430$, $A_q(0)=0.510(25)$, $D_g(0)=-1.275$
and $D_q(0)=-1.30(49)$. The parameters of the gluon GFFs are from the holographic model~\cite{Mamo:2022eui}, in overall agreement with those recently reported by the E12-007 collaboration~\cite{Duran:2022xag}.
The quark GFFs are obtained by recent Lattice results in~\cite{Hackett:2023rif}, as
illustrated in Fig.~\ref{fig:pressureQ}.
To fix  the sum rule 
$A_q(0)+A_g(0)=1$, we set $A_q(0)$=0.57, which is slightly larger than the lattice results. The remaining EMT form factor $B$ is null in dual gravity~\cite{Mamo:2022eui}, and is very 
small in unpolarized Lattice calculations~\cite{Hackett:2023rif} and global analysis~\cite{Guo:2023ahv}.

Although, the holographic EMT form factors are given in terms of hypergeometric functions~\cite{Mamo:2022eui},
Eq.~(\ref{ABFF}) provides a good approximation for a wide range of momenta. They are in agreement with the hard scattering rules asymptotically. We will not
consider the additional $\bar C_{q,g}$ form factors as they are absent in the holographic construction, and add to zero in physical observables.
Alternative discussions to some of these form factors can be found in~\cite{Kharzeev:2021qkd,Ji:2021mtz,Hatta:2021can,guo:2021ibg,Sun:2021gmi,Wang:2022vhr}.

In this work, we assume that the nucleons in the deuteron are quasi-free particle since the binding energy is very small. For on-shell nucleons, we can use the Gordon identity to rewrite Eq.~(\ref{A1tmp}) as Eq.~(\ref{A1}), which is more convenient for the non-relativistic reduction. More specifically, we have
\begin{widetext}
\be
\label{A1}
T_N^{\mu\nu}(p_2,p_1)=
\overline{u}(p_2)\bigg(
A(k)\frac{\bar{P}^\mu \bar{P}^\nu}{m_N}+ (A(k)+B(k))
\frac{i\bar{P}^{(\mu}\sigma^{\nu)\alpha}k_\alpha}{2m_N}
+C(k)\frac{k^\mu k^\nu-\eta^{\mu\nu}k^2}{m_N}\bigg)u(p_1)\,,
\ee

To probe the EMT in the deuteron at low and intermediate momentum transfers, we will use a non-relativistic reduction,
with the assumption that it holds for
$k/m_N$ of about 1. The justification 
for this assumption can only be made a posterioriti, by comparing to possibly future diffractive experiments. We recall 
that a similar assumption works reasonably well for the electromagnetic 
probes in the deuteron, at the nucleon mass scale~\cite{Riska:1989bh}. 
With this in mind, the non-relativistic reduction of Eq.~(\ref{A1}) reads

\be
\label{A3}
T_N^{00}(k) &=& \bigg(A(k)\,m_N
+\bigg(\frac 18 A(k)-\frac 14 B(k)+C(k)\bigg)\frac{\vec{k}^2}{m_N}\bigg)
+\bigg(\frac{1}{2}A(k)+B(k)\bigg)\frac{(\sigma\times ik)\cdot P}{2m_N}+{\cal O}\bigg(\frac {\vec{k}^3}{m^2_N}, \frac{\vec{P}^2}{m_N}\bigg),\nonumber\\
\nonumber\\
T_N^{0j} (k)&=&  (A(k)+B(k))\,\frac{(\sigma\times ik)^j}{4}
+A(k)\,P^j + {\cal O}\bigg(\frac {\vec{k}^3}{m^2_N}, \frac{\vec{P}^2}{m_N}\bigg),\nonumber\\
T_N^{jl}(k)&=&  (A(k)+B(k))
\frac {(\sigma\times ik)^{(j}P^{l)}}{2m_N}+{C(k)\frac{k^lk^j-\delta^{jl}\vec{k}^2}{m_N}}+{\cal O}\bigg(\frac {\vec{k}^3}{m^2_N}, \frac{\vec{P}^2}{m_N}\bigg),
\ee
\end{widetext}
where $P^j$ is the momentum of spectator shown in Fig~\ref{fig:BREIT} and only terms linear in  $P^j$ are retained. 
This additional assumption is justified 
in the analysis of the EMT of the deuteron to follow. Indeed, the higher order terms in the expansion when evaluating in a deuteron state,  are controlled by the binding energy $E_D=2.225\,{MeV}$ which is small,
$$\frac{\langle \vec{P}^2\rangle}{m_N^2}\approx  \frac{E_D}{m_N}\approx 10^{-3}.$$
With this in mind, and dropping the ${\cal O}$ notations for convenience, Eq.~(\ref{A3}) yields
\bea\label{eq:A3tmp}
T^{00}_N(k)&=&T_{M}(k)+T_{SP}(k)
\frac{(\sigma\times ik)\cdot P}{2m^2_N},\nonumber\\
T^{0j}_N(k)&=&T_{S}(k) 
\frac{(\sigma\times ik)^j}{4m_N}+A(k)P^j,\nonumber\\
T^{jl}_N(k)&=&T_S(k)
\frac {(\sigma\times ik)^{(j}P^{l)}}{2m_N^2}+{C_M(k)\frac{k^jk^l-\delta^{jl}\vec{k}^2}{m_N^2}}.\nonumber\\
\eea
For simplicity, we use the notation  $T_M(k)$, $T_S(k)$, $T_{SP}(k)$ to represent the contributions in Eq.~(\ref{A3}) for mass, spin, and spin-recoil. They can be related to form factor $A(k)$, $B(k)$ and $C(k)$ through,
\bea
T_{M}(k)&=&A(k)\,m_N
\nonumber\\
&+&\bigg(\frac 18 A(k)-\frac 14 B(k)+C(k)\bigg)\frac{\vec{k}^2}{m_N},
\nonumber\\
T_{S}(k)&=&m_N\,\left(A(k)+B(k)\right),  \nonumber\\
T_{SP}(k)&=&m_N\,\left(\frac 12 A(k)+B(k)\right),  \nonumber\\
C_{M}(k)&=&m_NC(k).  \nonumber\\
\eea
Since $T^{00}$ and $T^{0j}$ are related to the nucleon energy and spin, the factors $T_M$, $T_S$ represent the form factor of momentum fraction and angular momentum in the nucleon.

\section{Deuteron EMT in the impulse approximation}
\label{SEC4}
In the first approximation, we can treat the proton and neutron in the deuteron as quasi-free. In the impulse approximation, the EMT are the expectation values of Eq.~(\ref{A3}) in the deuteron state. Since the EMT is isoscalar,  the contributions of the proton and neutron add equally

\begin{figure}[htbp]
    \centering
\includegraphics[height=4cm,width=0.8\linewidth]{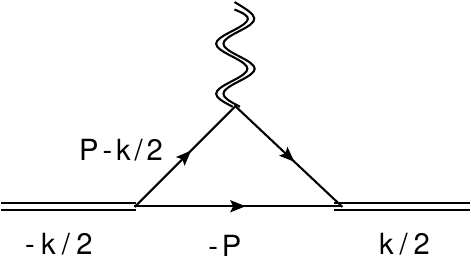}
    \caption{A graviton through 
    $T^{\mu\nu}(k)$, striking a nucleon in a deuteron, with a recoiling spectator of momentum $-P$.}
    \label{fig:BREIT}
\end{figure}

\begin{widetext}
\be
\label{D2X}
\big<+\frac k2 m'\big| T_N^{00}(k) \big|-\frac k2 m\big>&=& 
2T_M(k)
\big<+\frac k2 m'\big| {\bf 1}\, \big|-\frac k2 m\big>
+2T_{SP}(k)\big<+\frac k2 m'\big| 
\frac{P\cdot(S\times ik)}{2m^2_N}\big|-\frac k2 m\big>, 
\nonumber\\
\big<+\frac k2 m'\big| T_N^{0j}(k) \big|-\frac k2 m\big>&=&  
2T_S(k)\,\big<+\frac k2 m'\big| \frac{(S\times ik)^j}{4m_N}\big|-\frac k2 m\big>+2m_NA(k)\big<+\frac k2 m'\big|  \frac{P^j}{m_N}\big|-\frac k2 m\big>,
\nonumber\\
\big<+\frac k2 m'\big| T_N^{jl}(k) \big|-\frac k2 m\big>&=& 
 2T_S(k)\,
\big<+\frac k2 m'\big|\frac {(S\times ik)^{(j}P^{l)}}{2m^2_N}\big|-\frac k2 m\big>+{2C_M(k)\frac{k^jk^l-\delta^{jl}\vec{k}^2}{m_N^2}}\big<+\frac k2 m'\big| {\bf 1}\, \big|-\frac k2 m\big>.
\nonumber\\
\ee
\end{widetext}
Here $m,m'$ refer to the azimuthal quantum numbers in Eq.~(\ref{D1X}) (not to be confused with the reduced mass).
The matrix elements in the deuteron state can be simplified using symmetry arguments, and the conservation of the EMT. Their physical interpretation is best in the Breit (brick-wall) frame as illustrated in Fig.~\ref{fig:BREIT}, with $k^0=0$  and $\vec k\cdot \vec P=0$.

The simplest matrix elements to evaluate do not involve the total momentum $P$, they will be evaluated first, followed by single 
and double total momenta. For the latter, we will rely on a
wavefunction prescription to avoid issues of hermiticity.

\subsection{Matrix element of {\bf 1}}
The details of the matrix element of ${\bf 1}$ will be provided
in full to show how all matrix elements are evaluated. 
More specifically and following the kinematics depicted in Fig.~\ref{fig:BREIT}, the matrix element can be defined as~\cite{Gross:1965zz}
\begin{widetext}
\bea
\label{D3X}
\big<+\frac k2 m'\big| {\bf 1}\big|-\frac k2 m\big>=
\int d^3P\,\Phi^\dagger_{m'} \bigg(P+\frac k4\bigg)\,{\bf 1}
\Phi_{m} \bigg(P-\frac k4\bigg)
=\int d^3r\,e^{\frac i2 k\cdot r}\varphi_{m'}^\dagger(r)\,
{\bf 1}\, \varphi_m(r),
\eea
\end{widetext}
where we used the wavepacket form for the deuteron
in-out states
\bea\label{eq:wfpacket}
\Phi_m\big(P\pm \frac k4\big)=\int d^3r\,e^{-i(P\pm \frac 14 k)\cdot r}\,\varphi_m(r).
\eea
Inserting Eq.~(\ref{D1X}) into Eq.~(\ref{D3X}) gives
\bea
&&\big<+\frac k2 m'\big| {\bf 1}\big|-\frac k2 m\big>=\nonumber\\
&&C_E(k)\delta_{mm'}-
2C_Q(k)\langle m'|(S\cdot\hat k)^2-\frac 13 S^2|m\rangle\nonumber\\
\eea
with the deuteron total spin and angular momentum
\bea
&&\vec S=\frac 12 (\vec\sigma_p+\vec \sigma_n)=\vec\sigma,\nonumber\\
&&\vec L=\vec L_p+\vec L_n=
\vec r\times \vec P,
\eea
and the form factors
\bea
\label{D4X}
C_E(k)&=&\int_0^\infty dr (u^2+w^2)\,j_0\bigg(\frac {kr}2\bigg),\nonumber\\
\label{D5X}
C_Q(k)&=&\frac {3}{\sqrt 2}
\int_0^\infty dr  \bigg(uw-\frac{w^2}{2\sqrt 2}\bigg)\,j_2\bigg(\frac {kr}2\bigg).\nonumber\\
\eea
The deviation from spherical symmetry follows from the D-wave content of the deuteron wavefunction. 
We note that $C_E(0)=1$ as expected from the deuteron charge normalization, and that near the forward limit
\bea
C_Q(k)\approx \frac{Q^E_D}4\,\vec{k}^2,
\eea
with the quadrupole moment $Q^E_D\approx 0.31\rm fm^2$.

\subsection{Matrix element of $(S\times ik)^j$ }
Similarly, the spin contribution can be obtained by symmetry using the Wigner-Eckart theorem
\begin{widetext}
\bea
\big<+\frac k2 m'\big| (S\times ik)^j\big|-\frac k2 m\big> 
=
\int d^3r\,e^{\frac i2 k\cdot r}\varphi_{m'}^\dagger(r)\,
(S\times ik)^j\, \varphi_m(r)
=C_S(k)\langle m'| (S\times ik)^j|m\rangle,
\eea
with the result
\bea
C_S(k)&=&\int_0^\infty dr j_0\bigg(\frac{kr}2\bigg)
\bigg(u^2-\frac{w^2}{2}\bigg) 
+\frac{1}{\sqrt{2}}\int_0^\infty dr j_2\bigg(\frac{kr}2\bigg)\bigg(wu+\frac{w^2}{\sqrt{2}}\bigg),
\eea
\end{widetext}
or by direct computation, by 
specializing to $i=2$, $m=1$ and $m'=0$, 
and choosing $k=k\hat 3$. In the forward limit
\bea
C_S(0)=C_I(0)-4C_P(0)=1-\frac 32 p_D=1-\frac 32 \,6.53\%\nonumber\\
\eea
with $C_P(0)$ given in Eq.~(\ref{CP0}) below, and $p_D=6.53\%$ the percentage of D-admixture in the deuteron.

\subsection{Matrix element of $P^j$}\label{subsec:Pi}
The first recoil contribution  $P^j$
to the energy momentum tensor can be obtained as follows
\begin{widetext}
\bea
\label{PONE1}
\big<+\frac k2 m'\big| P^j\big|-\frac k2 m\big>
=
\int d^3{P}\,
\Phi^\dagger_{m'}(P+\frac 14 k)
 P^j\Phi_m(P-\frac 14 k)
=\frac i2\int d^3r
\,e^{\frac i2 k\cdot r}\,
\big(\partial_j\varphi_{m'}^\dagger(r)\varphi_m(r)-\varphi_{m'}^\dagger(r)\partial_j\varphi_m(r)\big).
\nonumber\\
\eea
Inserting the explicit derivative of the deuteron wavefunction
\bea
\label{PONE2}
\partial_j\varphi_m(r)&=&
\bigg(\left(\frac{u'(r)}{r}-\frac{u(r)}{r^2}-\frac{w(r)}{\sqrt{2}r^2}\right)\hat{r}^j
+\left(\frac{w'(r)}{\sqrt{8}r}-\frac{3w(r)}{\sqrt{8}r^2}\right)S_{12}(\hat{r})\hat{r}^j
+\frac{3w(r)}{\sqrt{8}r}\left(\frac{\sigma_1^j\sigma_2\cdot\hat{r}+\sigma_2^j\sigma_1\cdot\hat{r}}{r}\right)\bigg)\chi_m\nonumber\\
\eea
in Eq.~(\ref{PONE1}), and using the identities
\bea
\frac{j_1(\frac{kr}{2})}{kr}=\frac{j_0(\frac{kr}{2})+j_2(\frac{kr}{2})}{6},\qquad\qquad
\int d^3re^{i\frac{\vec{k}\cdot{\vec{r}}}{2}}\hat{r}^j=4\pi\int r^2dr\,j_1\bigg(\frac{kr}{2}\bigg)\,i{\hat k}^j
\eea
\end{widetext}
we can finally reduce Eq.~(\ref{PONE1}) to 
\bea
\label{PONE3}
\big<+\frac k2 m'\big| P^j\big|-\frac k2 m\big>=
C_P(k)\,\langle m'|({S}
\times i{k})^j|m\rangle
\eea
with
\bea
C_P(k)=
\int dr\frac{3w^2 }{8}\bigg(j_0\bigg(\frac{kr}{2}\bigg)+j_2\bigg(\frac{kr}{2}\bigg).\bigg)
\eea
Its forward contribution is readily tied to the admixture of D-state in the deuteron
\bea
\label{CP0}
C_P(0)=\frac 38\, p_D= \frac 38 \,6.53\%,
\eea
Eq.~(\ref{PONE3}) is manifestly transverse to the direction of momentum k. The Breit
frame projection through $P^i\rightarrow \tilde{P}^i$ as in Eq.~(\ref{SUB}) below, leaves it unchanged.

\subsection{Matrix element of 
$(S\times ik)^{(j}P^{l)}$}
This matrix element can be obtained by evaluating first 
\bea
\label{PISJ}
t^{jl}&=&\left< +\frac{k}{2}m'\bigg|
\frac 12\bigg(P^j(\vec{S}\times i\vec{k})^l+P^l(\vec{S}\times i\vec{k})^j\bigg)\bigg|-\frac{k}{2}m\right>.\nonumber\\
\eea

More specifically, the reduction of Eq.~(\ref{PISJ}) follows the same reasoning as above, with
\begin{widetext}
\bea\label{eq:maxSP}
t^{jl}&=&-\frac{1}{4}\int d^3r 
e^{\frac i2 k\cdot r}\left(\partial^j\varphi_m'^\dagger(r)(\vec{S}\times \vec{k})^l\varphi_m(r)-\varphi_m'^\dagger(r)(\vec{S}\times \vec{k})^l\partial^j\varphi_m(r)\right)+(l\leftrightarrow j)
\nonumber\\
&=&-\int d^3r 
e^{\frac i2 k\cdot r}
\left(\frac{\sqrt{2}r\,(u'w-w'u)-w^2+2\sqrt{2}uw}{16r^3}\right)(\hat{r}^j(\vec{S}\times\vec{k})^lS_{12}-\hat{r}^jS_{12}(\vec{S}\times\vec{k})^l)
\nonumber\\
&-&\int d^3r 
e^{\frac i2 k\cdot r}
\frac{3wu}{4\sqrt{8}r^3}\left((\sigma_1^j\sigma_2\cdot \hat{r}+\sigma_2^j\sigma_1\cdot \hat{r})(\vec{S}\times\vec{k})^l-(\vec{S}\times\vec{k})^l(\sigma_1^j\sigma_2\cdot \hat{r}+\sigma_2^j\sigma_1\cdot \hat{r})\right)
\nonumber\\
&-&\int d^3r 
e^{\frac i2 k\cdot r}
\frac{3w^2}{32r^3}\left((\sigma_1^j\sigma_2\cdot \hat{r}+\sigma_2^j\sigma_1\cdot \hat{r})(\vec{S}\times\vec{k})^lS_{12}-S_{12}(\vec{S}\times\vec{k})^l(\sigma_1^j\sigma_2\cdot \hat{r}+\sigma_2^j\sigma_1\cdot \hat{r})\right)
\nonumber\\
&+&(l\leftrightarrow j)
\nonumber\\
&=&\frac{1}{k}\int dr\left(\frac{\sqrt{2}r\,(u'w-w'u)-w^2+2\sqrt{2}uw}{16r}\right)\Bigg[\left(\frac{10}{kr}j_2\left(\frac{kr}{2}\right)-j_1\left(\frac{kr}{2}\right)\right)
\nonumber\\
&\times&\left(24\frac{k^jk^l}{\vec{k}^2}Q^{\alpha\beta}k_\alpha k_\beta
-12\left(Q^{j\beta}k_l k_\beta+Q^{l\beta}k_j k_\beta\right)\right)-\frac{48j_2\left(\frac{kr}{2}\right)}{kr}(\delta^{jl}Q^{\alpha\beta}k_\alpha k_\beta-\vec{k}^2Q^{jl})\Bigg]
\nonumber\\
&+&\frac{1}{k}\int dr\left(\frac{wu}{\sqrt{8}r}-\frac{w^2}{8r}\right)j_1\left(\frac{kr}{2}\right)
\left(6\delta^{jl}Q^{\alpha\beta}k_\alpha k_\beta-6\vec{k}^2Q^{jl}\right)
\nonumber\\
&+&\frac{1}{k}\int dr\frac{9w^2}{8r}j_1\left(\frac{kr}{2}\right)
\left((\vec{S}\times k)^l (\vec{S}\times k)^j+(\vec{S}\times k)^j(\vec{S}\times k)^l\right)
\nonumber\\
&+&\frac{1}{k}\int dr\frac{3w^2}{2r}\frac{j_2\left(\frac{kr}{2}\right)}{kr}
\Bigg(4(\delta^{jl}\vec{k}^2-k^lk^j)-3(\vec{S}\times k)^l (\vec{S}\times k)^j-3(\vec{S}\times k)^j(\vec{S}\times k)^l
\nonumber\\
&+&3(2\delta^{jl}Q^{\alpha\beta}k_\alpha k_\beta-Q^{j\beta}k_l k_\beta-Q^{l\beta}k_j k_\beta)\Bigg).\nonumber\\
\eea

The matrix element in Eq.~(\ref{eq:maxSP}) lacks manifest transversality, i.e., $k^lt^{jl}\neq 0$.  This 
is caused by the off-shell character of the struck nucleon which leads to the violation of the kinematic condition: $\vec{k}\cdot\vec{P}\neq 0$ say in the Breit frame.
To enforce the Breit frame condition $\vec k\cdot\vec P=0$ in the matrix elements, we will make the operator substitution
\bea
\label{SUB}
P^\mu\rightarrow \tilde{P}^\mu=P^\mu-\frac {(k\cdot P)}{k^2}\,k^\mu\, 
\eea
This replacement is carried out in the evaluation of the EMT nucleon matrix elements, rather than the evaluation of the wave function in Eq.~(\ref{eq:wfpacket}),
followed by the Breit frame substitution in Eq~(\ref{SUB}). This  amounts to the projection
\begin{eqnarray}
\label{TIJPROJ}
\tilde{t}^{jl}=t^{jl}-\frac{k^lt^{j\alpha}k^\alpha+k^jt^{l\alpha}k^\alpha}{k^2}+k^jk^l\frac{k^\alpha t^{\alpha\beta}k^\beta}{k^4}    
\end{eqnarray}
 when evaluating the matrix elements of the EMT in the deuteron state.  The upshot of this substitution, is the manifest conservation of the the recoil corrections in the deuteron EMT,
\begin{eqnarray}
\tilde{t}^{jl}&=&\frac{k^lk^j-\delta^{jl}\vec{k}^2}{2}D^{SP}_0+(k^jk^\alpha Q^{l\alpha}+k^lk^\alpha Q^{j\alpha}-\vec{k}^2Q^{jl}-\delta^{jl}Q^{\alpha\beta}k_\alpha k_\beta)D^{SP}_2
\nonumber\\
&+&(k^jk^l-\delta^{jl}\vec{k}^2)Q^{\alpha\beta}\hat k_\alpha \hat k_\beta D^{SP}_3
\end{eqnarray}
with
\begin{eqnarray}
D^{SP}_0&=&-\int dr\,\frac{3 w^2 j_1\left(\frac{k r}{2}\right)}{k  r},
\nonumber\\
D^{SP}_2&=&+\int dr\,\frac{3 k r w \left(\sqrt{2} u+w\right) j_1\left(\frac{k r}{2}\right)-6 j_2\left(\frac{k r}{2}\right) \left(\sqrt{2} u (2 w-r w')+w \left(\sqrt{2} r u'+2 w\right)\right)}{2 \vec{k}^2  r^2},
\nonumber\\
D^{SP}_3&=&-\int dr\,\frac{3 \left(2 j_2\left(\frac{k r}{2}\right) \left(2 \sqrt{2} u (r w'-2 w)+w \left(5 w-2 \sqrt{2} r u'\right)\right)+k r w \left(2 \sqrt{2} u-w\right) j_1\left(\frac{k r}{2}\right)\right)}{2\vec{k}^2 r^2}.
\end{eqnarray}
The net result is the matrix entry
\bea
&&\big<+\frac k2 m'\big|{(S\times ik)^{(j}P^{l)}}\big|-\frac k2 m\big> =
\frac{(k^jk^l-\delta^{jl}\vec{k}^2)}{2}D^{SP}_0\,\delta_{mm'}\nonumber\\
&&+\langle m'|(k^jk^\alpha Q^{l\alpha}+k^lk^\alpha Q^{j\alpha}-k^2Q^{jl}-\delta^{jl}Q^{\alpha\beta}k_\alpha k_\beta)|m\rangle D^{SP}_2
+(k^jk^l-\delta^{jl}\vec{k}^2)
\hat k_\alpha \hat k_\beta \,
\langle m'|Q^{\alpha\beta}|m\rangle\,D^{SP}_3
\eea
\end{widetext}
which is manifestly transverse, with all invariant form factors $D^{SP}_0, D^{SP}_2, D^{SP}_3$ 
finite in the forward limit.

\begin{figure*}[htbp] 
\begin{minipage}[b]{.45\linewidth}
\hspace*{-0.3cm}\includegraphics[width=1.1\textwidth, height=5.5cm]{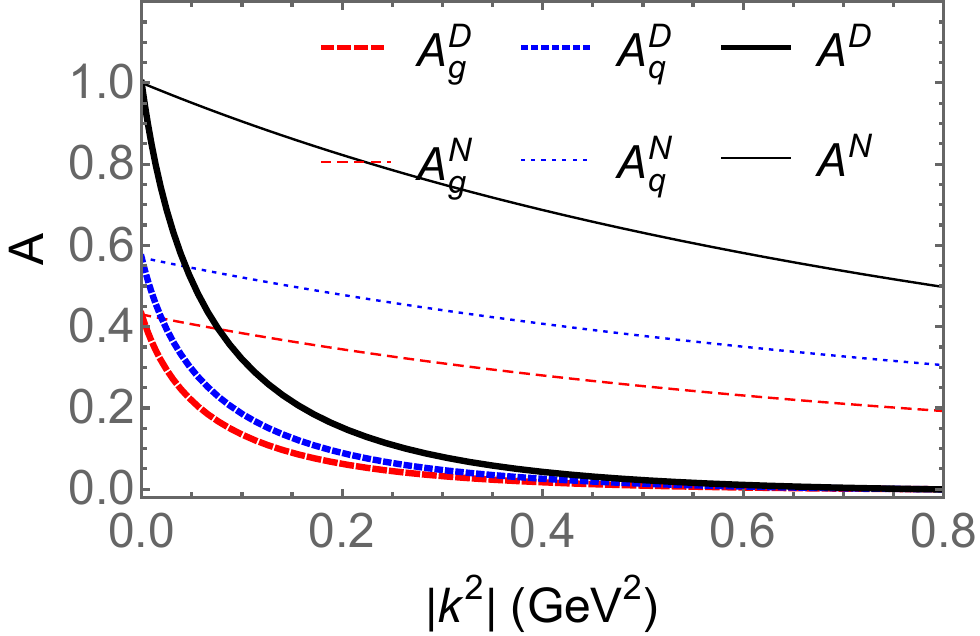}   
 \vspace{0pt}
\end{minipage}
\hfill
\begin{minipage}[b]{.45\linewidth}   
\hspace*{-0.55cm} \includegraphics[width=1.1\textwidth, height=5.5cm]{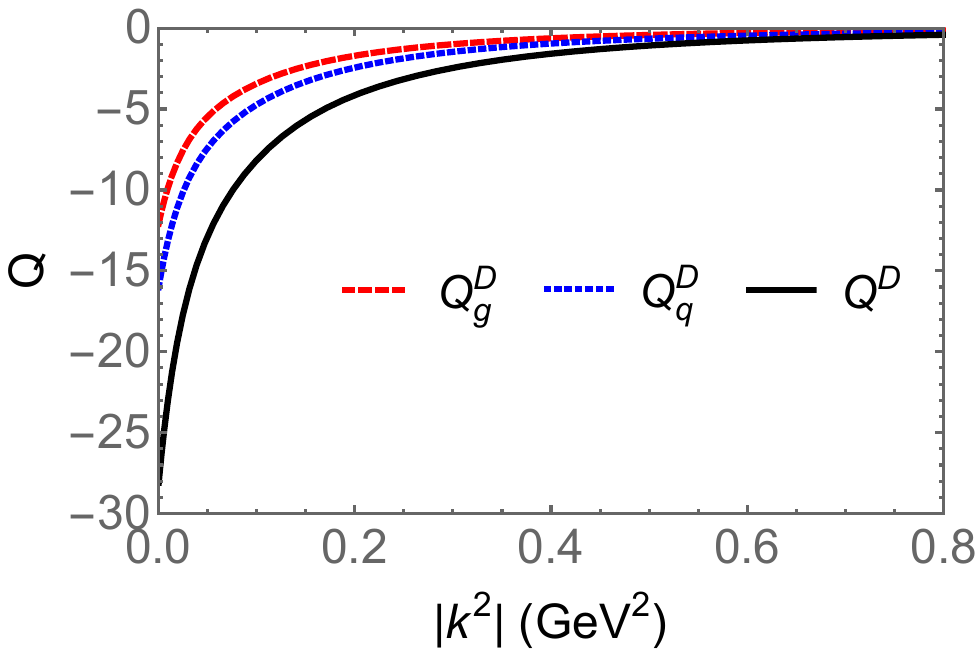} 
   \vspace{0pt}
\end{minipage}  
\\[-0.4cm]
\begin{minipage}[t]{.45\linewidth}
\hspace*{-0.5cm}\includegraphics[width=1.12\textwidth, height=5.5cm]{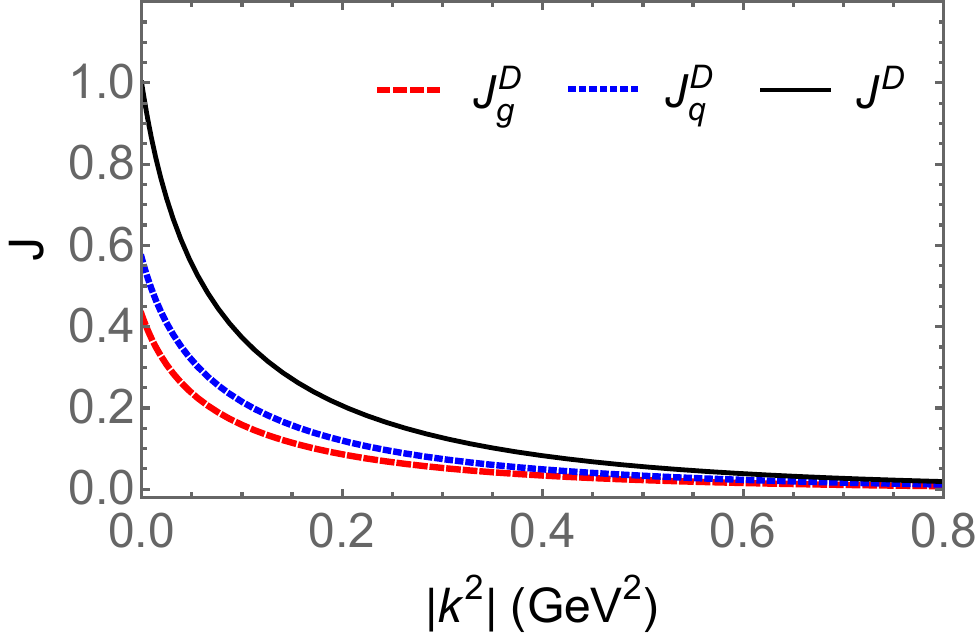}   
 \vspace{0pt}
\end{minipage}
\hfill
\begin{minipage}[t]{.45\linewidth}   
\hspace*{-0.85cm} \includegraphics[width=1.1\textwidth, height=5.5cm]{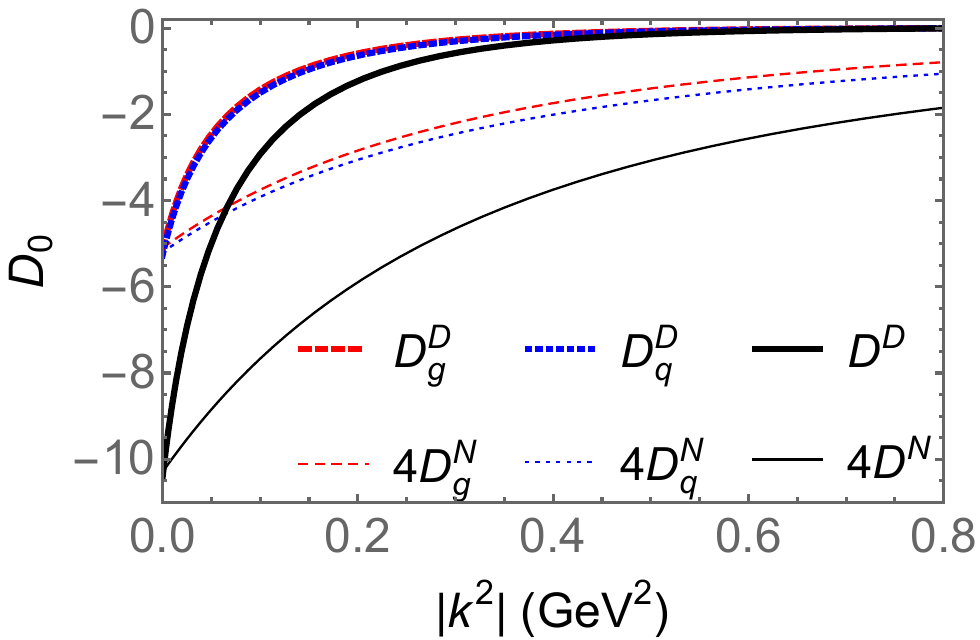}  
   \vspace{0pt}
\end{minipage} 
\begin{minipage}[t]{.45\linewidth}
\hspace*{-0.5cm}\includegraphics[width=1.12\textwidth, height=5.5cm]{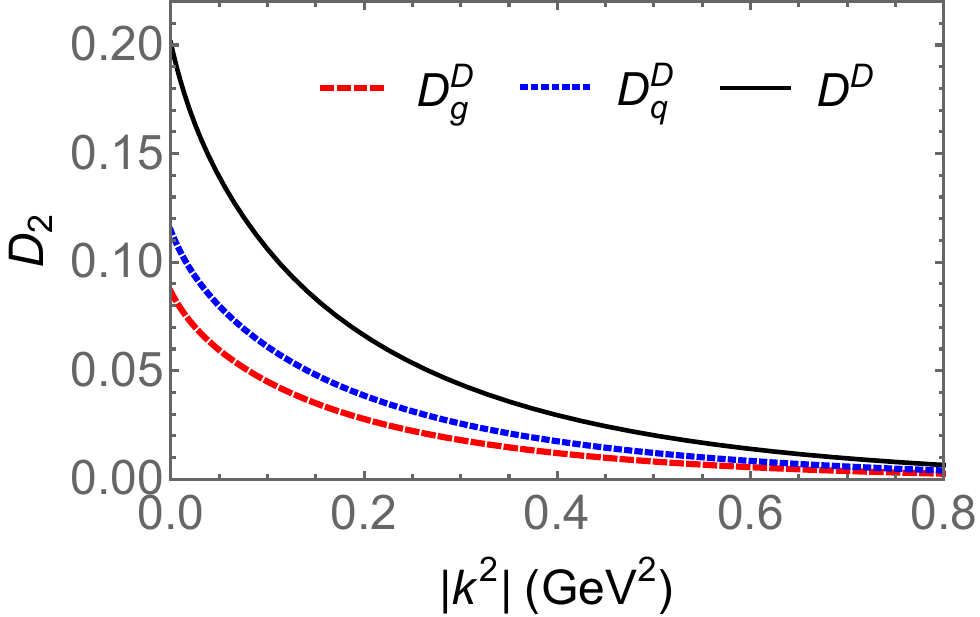}   
 \vspace{0pt}
\end{minipage}
\hfill
\begin{minipage}[t]{.45\linewidth}   
\hspace*{-0.85cm} \includegraphics[width=1.1\textwidth, height=5.5cm]{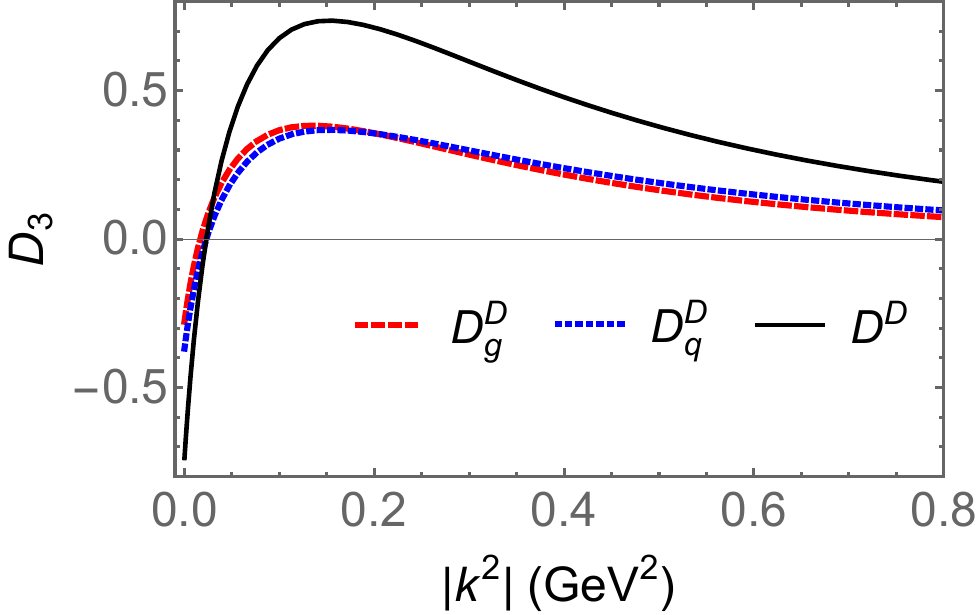}  
   \vspace{-12pt}
\end{minipage}
\caption{The deuteron  invariant EMT form factors in Eq.~(\ref{EMTDEUTIMP}) in the impulse approximation: gluon (red-dashed), quark (blue-dotted) and gluon+quark (black-solid),
compared to the nucleon  $A^N$ and $D^N$
(thinner dashed, dotted and solid).}
\label{fig:imp_result}
\end{figure*}

\subsection{Summary EMT in impulse approximation}
In summary, the non-relativistic contributions to the deuteron EMT in the impulse approximation and in linear order in the recoil momentum of the spectator nucleon, are given by
\begin{widetext}
\bea
\label{DR1X}
T_D^{00}(k, m', m)=&&2T_M(k)\bigg(C_E(k)\delta_{mm'}-2C_Q(k)\hat k_\alpha \hat k_\beta \langle m'|Q^{\alpha\beta}|m\rangle\bigg)\nonumber\\
&+&2T_{SP}(k)
\bigg(-D^{SP}_0(k)\delta_{mm'}-\left(D^{SP}_2(k)+2D^{SP}_3(k)\right)\,
\hat k_\alpha \hat k_\beta \langle m'|Q^{\alpha\beta}|m\rangle\bigg)\,\frac {\vec{k}^2}{2m^2_N}\nonumber\\
=&&m_DA^D(k)\delta_{mm'}+Q^D (k)\frac{k_\alpha k_\beta}{2m_D} \langle m'|Q^{\alpha\beta}|m\rangle,
\nonumber\\
T_D^{0j}(k, m', m)=&&
2T_S(k)\, C_S(k)\,\langle m'|\frac{(S\times ik)^j}{4m_N}|m\rangle+ 2m_NA(k)\,C_P(k)\,\langle m'|\frac{(S\times ik)^j}{m_N}|m\rangle
\nonumber\\
=&&J^D(k)\frac{\langle m'|(\vec{S}\times i\vec{k})^j|m\rangle }{2},
\nonumber\\
T^{jl}_D(k, m', m)=&& 2T_S(k)\,
\bigg(\frac{(k^jk^l-\delta^{jl}\,\vec{k}^2)}{2}D^{SP}_0(k)\delta_{mm'}+(k^jk^l-\delta^{jl}\vec{k}^2)Q^{\alpha\beta}\hat k_\alpha \hat k_\beta D^{SP}_3(k)\nonumber\\
&+&
\langle m'|(k^jk^\alpha Q^{l\alpha}+k^lk^\alpha Q^{j\alpha}-\vec{k}^2Q^{jl}-\delta^{jl}Q^{\alpha\beta}k_\alpha k_\beta)|m\rangle \, D^{SP}_2(k)\bigg)\, \frac 1{2m^2_N}
\nonumber \\
&+&{2C_M(k)\frac{C_E(k)(k^jk^l-\delta^{jl}\vec{k}^2)\delta_{mm'}-2C_Q(k)(k^jk^l-\delta^{jl}\vec{k}^2)\hat k_\alpha \hat k_\beta \langle m'|Q^{\alpha\beta}|m\rangle}{m_N^2}}
\nonumber \\
=&&D_0^D(k)\frac{k^jk^l-\delta^{jl}\vec{k}^2}{4m_D}\delta_{m'm}
+D_3^D(k)\frac{(k^jk^l-\delta^{jl}\vec{k}^2)\hat k_\alpha \hat k_\beta \langle m'|Q^{\alpha\beta}|m\rangle}{4m_D}
\nonumber\\
&+&D_2^D(k)\frac{\langle m'|(k^jk^\alpha Q^{l\alpha}+k^lk^\alpha Q^{j\alpha}-\vec{k}^2Q^{jl}-\delta^{jl}Q^{\alpha\beta}k_\alpha k_\beta)|m\rangle }{2m_D}.
\eea
\end{widetext}
Our conventions for the deuteron EMT invariant form factors, follow the general
spin-1 conventions introduced in~\cite{Polyakov:2019lbq}.
In the impulse approximation and
to linear order in the recoil momentum 
of the spectator nucleon (in short hand notations), they are
\bea
\label{EMTDEUTIMP}
A^D&=&\frac{2}{m_D}\left(T_MC_E-\frac{\vec{k}^2}{2m_N^2}
T_{SP}D^{SP}_0\right),\nonumber\\
Q^D&=&-\frac{2m_D}{\vec{k}^2}\left(4T_MC_Q
+\frac{\vec{k}^2}{m^2_N}T_{SP}(D^{SP}_2+2D^{SP}_3)\right),\nonumber\\
J^D&=& \frac{T_SC_S}{m_N}+4AC_P,\nonumber\\
D^D_0&=&\frac{4m_D}{m_N^2}
\bigg(\frac12 T_S D^{SP}_0+2C_MC_E\bigg),\nonumber\\
D_2^D&=&\frac{2m_D}{m_N^2}T_SD^{SP}_2,\nonumber\\
D_3^D&=&\frac{4m_D}{m_N^2}
\big(T_SD^{SP}_3-4C_MC_Q)
\eea
with the deuteron quadrupole form factor
\bea
Q(k)=-\frac {4m_D^2}{\vec{k}^2}C_Q(k)\rightarrow Q_D
\eea
that reduces to the deuteron quadrupole moment in the forward direction.

The numerical results for low and intermediate momenta $k\leq m_N$, are shown in Fig~\ref{fig:imp_result}. Note that
the quark and gluon contributions to $D_0$ and $D_3$ are comparable, since  $C_g(k)$ and $C_q(k)$ are very similar in Eq.~(\ref{ABFF}). In the impulse approximation, the spin averaged deuteron D-value at the origin is $D_0(0)=-10.43$, which  is to be compared to $D_0(0)=-13.126$ in the Skyrme model~\cite{GarciaMartin-Caro:2023klo}, and 
$D_0(0)=-24.33$ in the relativistic light cone convolution model~\cite{Freese:2022yur}.

For $B(k)\approx 0$ and at low momenta
$k\ll m_N$, we have
$$T_M(k)\approx T_S(k)\approx 2T_{SP}(k)\approx m_NA(k).$$
The deuteron invariant EMT form factors (\ref{EMTDEUTIMP}) simplify (short hand notation)
\bea
\label{DEUTAPPROX}
A^D&\approx& A(k)C_E,\nonumber\\
Q^D&\approx& -\frac {4m_D^2}{\vec{k}^2}A(k) C_Q-2A(k)(D^{SP}_2+2D^{SP}_3),\nonumber\\
J^D&\approx& A(k) (C_S+4C_P),
\eea
for the mass $A^D$, quadrupole $Q^Q$ and
momentum $J^D$ respectively. For the deuteron D-terms,
we have (short hand notation)
\bea
D_0^D&\approx& 4A(k)D^{SP}_0+16C(k)C_E,\nonumber\\
D_2^D&\approx& 4A(k)D^{SP}_2,\nonumber\\
D_3^D&\approx& 8A(k)D^{SP}_3-32C(k)C_Q,
\eea
for the standard tensor $D_0^D$, tensor spin-spin
$D_2^D$ and tensor-quadrupole $D_3^D$, respectively. 

The three deuteron D-form factors  in the impulse approximation, can be used to describe the
spatial distributions of the pressure and shear force inside the deuteron as probed by a graviton or
a graviton-like probes. Using the conventions for the pressure and shear introduced in~\cite{Polyakov:2002yz}, we show in Fig.~\ref{fig:pressure_D} their distribution
inside the deuteron. The formulas are put in Appendix~\ref{app:pre}.
We have separated the quark and gluon contributions, following their 
separation in the nucleon form factors~(\ref{ABFF}). Here $p_{\{0,2,3\},\{g,q\}}$ refer to the pressure distributions carried by the quarks and gluons separately, and $s_{\{0,2,3\},\{g,q\}}$ refer
to the shear distributions carried by the quarks and gluons also separately. We note that the pressure distributions and shear forces obtained in this work satisfied the von Laue conditions mentioned in~\cite{Polyakov:2019lbq}. The sign of $p_{0,g}$ changes around $r=1.4$ fm, which is further than the one reported in the nucleon in~\cite{Mamo:2022eui}. Both $p_{0,g}$ and $s_{0,g}$ have longer tails in  comparison to the nucleon.

\begin{figure}
\centering
    \includegraphics[height=5cm,width=0.9\linewidth]{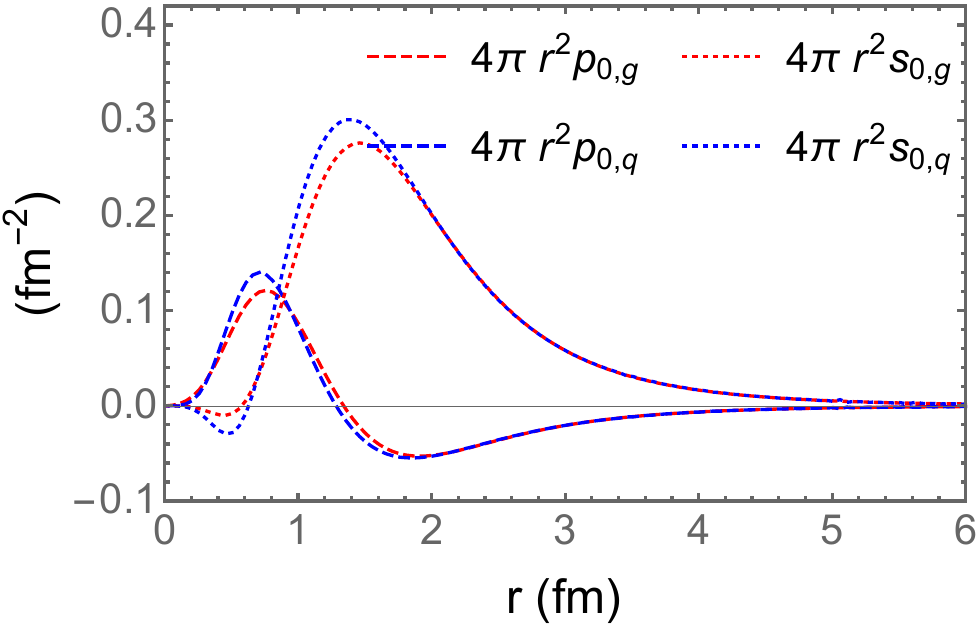}
    \includegraphics[height=5cm,width=0.9\linewidth]{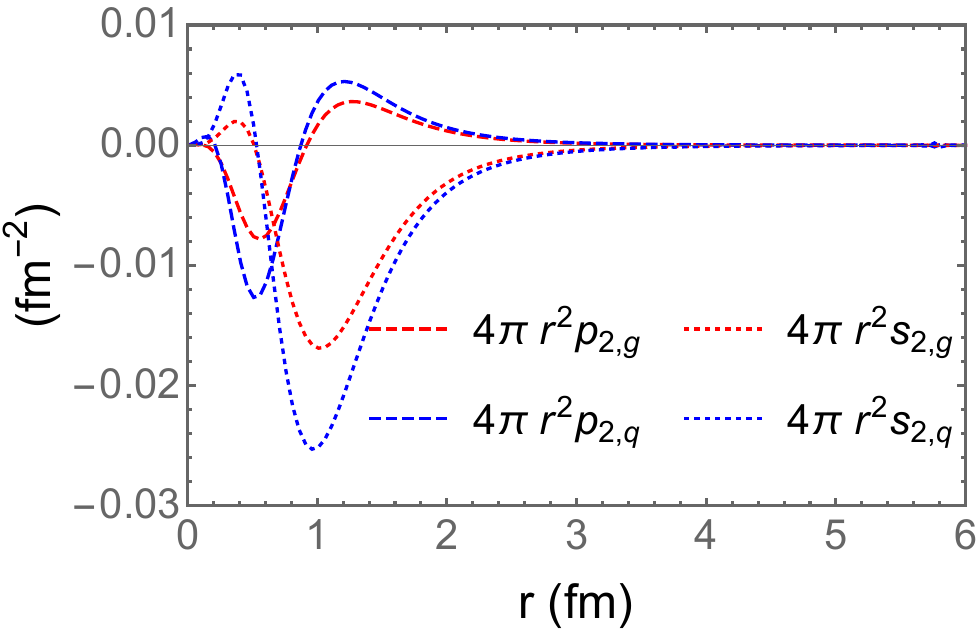}
    \includegraphics[height=5cm,width=0.9\linewidth]{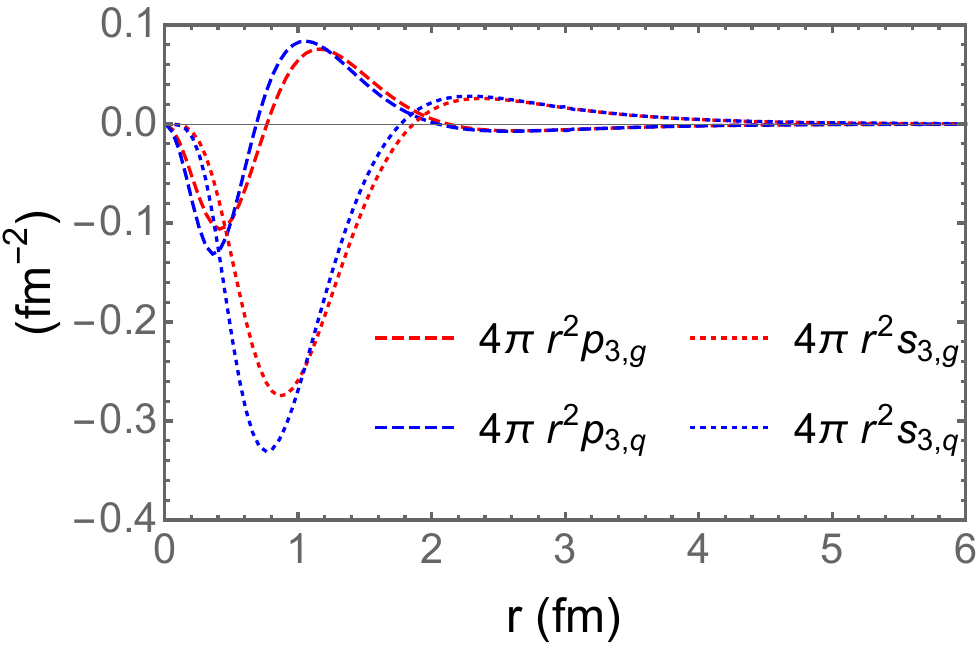}
    \caption{The gluon and quark pressure $p_{0,2,3}$ and shear $s_{0,2,3}$ distributions  inside  the deuteron, in the impulse approximation.}
    \label{fig:pressure_D}
\end{figure}

\section{Helium-4 EMT in the impulse approximation}
\label{SEC5}
We will start with the simplest $0^{++}$ helium-4 nucleus,  a scalar particle both in spin and isospin. The ground state of 
helium-4 is composed of 2 protons and 2 neutrons in a purely S-wave. 
Its $0^{++}$ EMT is characterized by two invariant form factors~\cite{Pagels:1966zza}
\bea
\label{T00++}
\langle p_2|T^{\mu\nu}| p_1\rangle&=&
\frac{P^\mu P^\nu}{m_\alpha} A^H(k)
\nonumber\\
&+&\frac{1}{4m_\alpha} (k^\mu k^\nu-g^{\mu\nu}k^2) D^H(k)\nonumber\\
\eea
In the impulse approximation, most of the EMT results for helium-4, can be inferred from those of the deuteron presented above with much simplifications. 
The same observations can be
extended to the lighter $0^{++}$ magic
nuclei.

\subsection{Helium-4 state}
To construct the helium-4 state, we will use the K-harmonics method to factor out the spurious center of mass motion~\cite{Badalian:1966wm,de1979first}. The method works well for few particle systems, when the multi-dimensional Schrodinger equation can be reduced to a one-dimensional hyper-radial distance times the lowest K-harmonics. 

The K-harmonics method becomes
increasingly involved for heavier nuclei, where the mean-field single particle approximation is more appropriate. However, the removal of the spurious center of mass motion is more challenging in the mean-field approach.
We will present both methods, when addressing helium-4 for comparison.

\subsubsection{K-harmonics method}
The ground state of helium-4 (alpha particle) is spin-isospin singlet, and reads
\bea
\label{ALPHA1}
&&\Phi_H[1, ..., 4]=\varphi[r_1,r_2,r_3,r_4]\,{\bf P}[\sigma(i),\tau(i)],
\nonumber\\
\eea
where ${\bf P}[]$ refers to the properly symmetrized spin-isospin wavefunction. In general, the ground state contains a smaller D-wave admixture~\cite{Santos:1982zz}, that we have ignored for simplicity.

To remove the spurious center of mass motion in Eq.~(\ref{ALPHA1}) using the K-harmonics method,  the pertinent Jacobi coordinates are~\cite{de1979first}
\bea
\vec \xi_1&=&\frac 1{\sqrt 2}(\vec r_2-\vec r_1),\nonumber\\
\vec \xi_2&=&\frac 1{\sqrt 6}
(\vec r_1+\vec r_2-2\vec r_3),\nonumber\\
\vec \xi_3&=&\frac 1{2\sqrt 3}(\vec r_1+\vec r_2+\vec r_3-3\vec r_4),\nonumber\\
\vec R_C&=&\frac 14 
(\vec r_1+\vec r_2+\vec r_3+\vec r_4),
\eea
the radial hyperdistance is
\bea
R^2= \frac 14 \sum_{i\neq j}
(\vec r_i-\vec r_j)^2=\vec \xi_1^2+\vec \xi_2^2+\vec \xi_3^2
\eea
and the center of mass motion factors out of the 4-particle Kinetic contribution 
\bea
\mathbb K=-\sum_{i=1}^4\frac {\nabla_i^2}{2m_N}\rightarrow -\frac 1{2m_N}\bigg(\frac{d^2}{dR^2}+\frac 8R \frac {d}{dR}-\frac{K_N^2}{R^2}\bigg) \nonumber\\
\eea

The hyper-spherical harmonics (HHs) are the eigenstates of the grand-angular momentum~\cite{de1979first}
\be
K_N^2\,{\cal Y}_{[K]}^{KLM_L}(\Omega_{\tilde N})=(K(K+3N-2))\,{\cal Y}_{[K]}^{KLM_L}(\Omega_{\tilde N})\nonumber\\
\ee
for atomic number $A$, with $N=A-1$. 
The $\tilde N=3N-1$ angles are fixed by the hyperspherical  symmetric Jacobi coordinates through
\bea\label{eq:Jcobicoor}
\xi_1&=&R\,\rm cos\theta(sin\theta_1cos\phi_1,sin\theta_1sin\phi_1,cos\theta_1),
\nonumber\\
\xi_2&=&R\,\rm sin\theta\cos\phi(sin\theta_2cos\phi_2,sin\theta_2sin\phi_2,cos\theta_2),
\nonumber\\
\xi_3&=&R\,\rm sin\theta sin\phi(sin\theta_3cos\phi_3,sin\theta_3sin\phi_3,cos\theta_3),\nonumber\\
\eea
They are valued as  $\theta_i\in[0,\pi]$, $\phi_i\in[0,2\pi]$, $\theta\in[0,\frac{\pi}{2}]$ and $\phi\in[0,\frac{\pi}{2}]$, 
with the angular volumes
\bea
&&\Omega_9=\int_0^{\pi/2} d\phi\,{\rm sin}^2\phi{\rm cos}^2\phi\int_0^{\pi/2}d\theta\,{\rm sin}^5\theta{\rm cos}^2\theta
\nonumber\\
&&\prod_{i=1}^3\,\int _0^{\pi}d\theta_i\int _0^{2\pi}d\phi_i{\rm sin}\theta_i=\frac{32\pi^4}{105}.\nonumber\\
\eea

The specific form of the HHs follows by recoupling the individual angular momenta $L_i$.  They are normalized as
\be
\int d\Omega_{\tilde N}\,{\cal Y}_{[K]}^{KLM_L\ *}(\Omega_{\tilde N})\,{\cal Y}_{[K']}^{K'L'M'_L}(\Omega_{\tilde N})
=\delta_{[K],[K']}
\ee
and their total number  is
\be
d_K=(2K+3N-2)\frac{(K+3N-3)!}{K!(3N-2)!}
\ee
For helium-4 with $A=4$ and  $N=3$, 
the $K=0$ HH has degeneracy $d_0=1$, and the $K=1$ HHs have degeneracy $d_1=9$. For $K=0$ case, the spin-isospin wave function can be written as~\cite{Castilho:1974}
\begin{widetext}
\bea
\label{eq:spispwf}
{\bf P}[\sigma(i),\tau(i)]&=&\frac{\sqrt{105}}{8\pi^2}
\bigg(([\sigma(1),\sigma(2)]_1[\sigma(3),\sigma(4)]_1)_{00}
([\tau(1),\tau(2)]_0[\tau(3),\tau(4)]_0)_{00}\nonumber\\
&&-
([\sigma(1),\sigma(2)]_0[\sigma(3),\sigma(4)]_0)_{00}
([\tau(1),\tau(2)]_1[\tau(3),\tau(4)]_1)_{00}
\bigg).
\eea
\end{widetext}
Here $\sigma(i), \tau(i)$ refer to the spin-isospin of the i-th nucleons. The subscripts refer to their recoupling to a total spin-isospin.

The general form of Eq.~(\ref{ALPHA1})  in hyper-spherical form modulo the spin factors, is
\bea
\varphi_{[K]}(R){\cal Y}_{[K]}^{KLM_L}(\Omega_{\tilde 8}),
\eea
with the S-wave solution for helium-4
\bea
\label{PHI4X}
\varphi_{[0]}(R){\cal Y}_{[0]}^{000}(\Omega_{\tilde 8})=\frac{\varphi(R)}{\sqrt{\Omega_9}}.
\eea
To eliminate the linear derivative in the hyperdistance in the Schrodinger equation, we will seek the radial wavefunction
\bea
\label{PHI4}
\varphi(R)=\frac{u(R)}{R^4}
\eea
with the reduced wavefunction satisfying
\bea\label{eq:sch_Hel_new}
u{''}-\frac {12}{R^2} u-\frac{2m_N}{\hbar^2}
(W(R)+V_C(R)-E)u=0,\nonumber\\
\eea
subject to the normalization
\bea
\int_0^\infty dR \,|u(R)|^2=1.
\eea
A large centrifugation emerges following the reduction to the hyperdistance.
Here $W(R)$ is the projection of the pair potential $V(r_{ij})$ on the $K=0$ harmonic, which can be obtained through~\cite{Castilho:1974}
\bea
W(R)=\frac{1}{\Omega_9}\int d\Omega_9\sum_{i<j}V(r_{ij}).
\eea
Since the helium-4 wavefunction is symmetric under the spatial exchange of any pair of nucleons, it follows that 
\bea
W(R)=\frac{6}{\Omega_9}\int d\Omega_9 V(\sqrt{2}R{\rm cos\theta}),
\eea
or equivalently, 
\bea
\label{eq:WR}
W(R)=\frac{315}4\int_0^1\,dx\,(1-x^2)^2x^2\,
V(\sqrt 2R x).
\eea
For the pair Coulomb potential, 
\bea
V_C(r_{ij})=\left(\frac{1}{2}+\tau_z(i)\right)\left(\frac{1}{2}+\tau_z(j)\right)\frac{e^2}{4\pi\,r_{ij}},
\eea
using the spin-isospin helium-4 wavefunction (\ref{eq:spispwf}), the  Coulomb potential can be reduced to
\begin{widetext}
\bea\label{eq:VC}
V_C(R)=\sum_{i<j}\int d\Omega_9{\bf P}^+(\sigma(i),\tau(i))V_C(r_{ij}){\bf P}(\sigma(i),\tau(i))
=\frac{35}{16\sqrt{2}R}\frac{e^2}{4\pi}=\frac{2.23{\rm MeV} {\rm fm}}{R}.
\eea
\end{widetext}
We note the recent applications of this method to 
the clustering of light nuclei in heavy-ion collisions~\cite{Shuryak:2019ikv}, and the charmed tetraquark states in~\cite{Miesch:2023hjt}.

\begin{figure}
\centering
    \includegraphics[height=5cm,width=0.9\linewidth]{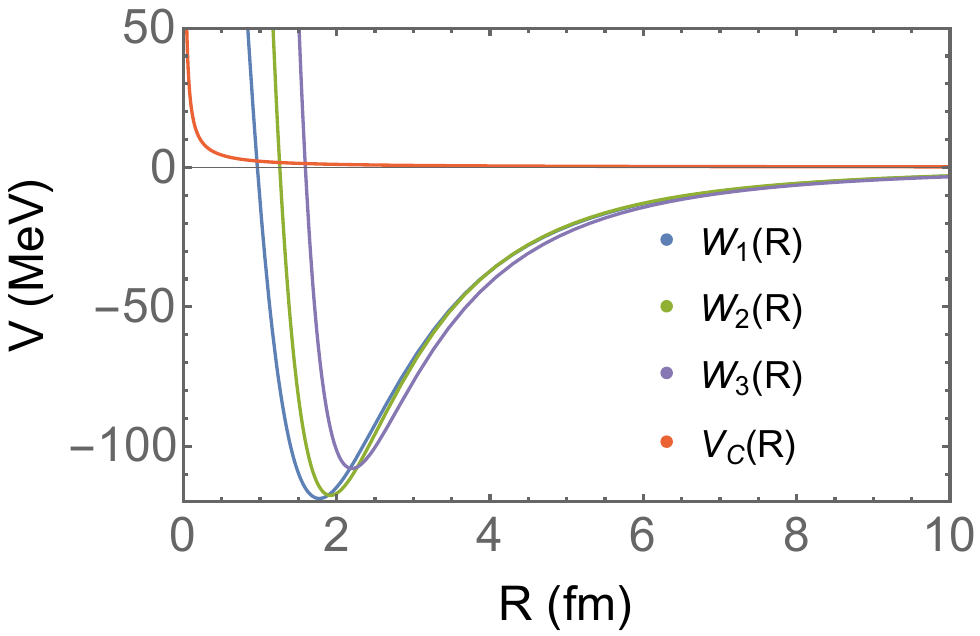}
    \includegraphics[height=5cm,width=0.9\linewidth]{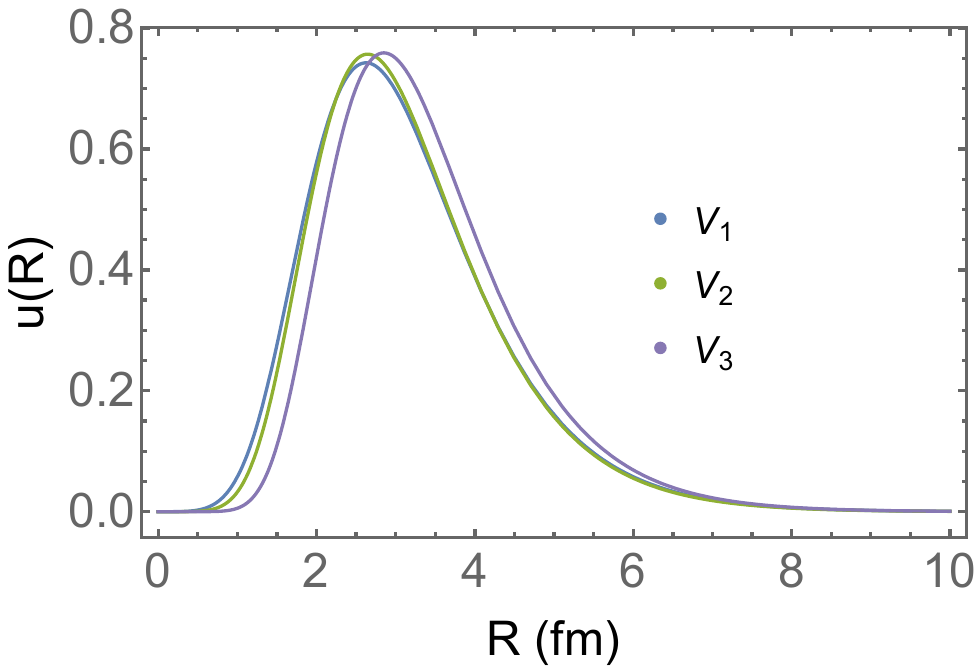}
    \caption{Top: The potential $W(R)$ and $V_C(R)$ defined in Eqs.~(\ref{eq:WR}) and ~(\ref{eq:VC}). $W_1$, $W_2$ and $W_3$ are obtained using $V_1$, $V_2$ and $V_3$ defined in Eqs.~(\ref{CAST}) and (\ref{eq:V3});
        Bottom:Helium-4 S-wave reduced wavefunction solution with potential $V_1$, $V_2$ and $V_3$ versus the hyperdistance.}
    \label{fig:WS4}
\end{figure}

Specific choices of the pair potential in Eq.~(\ref{eq:WR}) for helium-4, were discussed in~\cite{Volkov:1965zz,Brink:1967zz,Castilho:1974}
\bea
\label{CAST}
V_1(r)&=&+144.86\,e^{-(r/0.82)^2}-83.34\,e^{-(r/1.6)^2},\nonumber\\
V_2(r)&=&+389.5\,e^{-(r/0.7)^2}-140.6 \,e^{-(r/1.4)^2},
\eea
with the potential energy in MeV units, and the spatial range in fm. The above potentials are obtained by studying the binding energy of various light nuclei~\cite{Volkov:1965zz,Brink:1967zz}. We have checked that the 
effect of the Coulomb interaction is negligible. The potentials in Eq.~(\ref{CAST}) capture schematically the repulsion at short distance (omega-exchange) and the attraction at large distance (pion-exchange).  They reproduce the binding energy, electromagnetic radius, and the electromagnetic form factor of helium-4 up to momenta of the order of $\frac 12 m_N$ as detailed in Appendix~\ref{APPA}.  This range can be extended to $\frac 23 m_N$ with our choice
\bea\label{eq:V3}
V_3(r)&=&+1310.21\,e^{-(r/0.7)^2}-467.97\,e^{-(r/1.16)^2}.
\nonumber\\
\eea 
The reduced S-wave solution to Eq.~(\ref{eq:sch_Hel_new}) for $V_1$, $V_2$ and $V_3$ are shown in Fig.~\ref{fig:WS4} versus the hyperdistance. The binding energy of helium-4 with potential $V_1$, $V_2$ and $V_3$ are -27.75\,${\rm MeV}$, -28.47\,${\rm MeV}$ and -29.3 ${\rm MeV}$, respectively. The large induced centrifugation by projection on the hyperdistance, causes it to peak at 
$2.5\,{\rm fm}$. In the later sections, we will present the GFFs of helium-4 obtained with potential $V_3$.

\begin{figure}
\centering
 \includegraphics[height=5cm,width=0.9\linewidth]{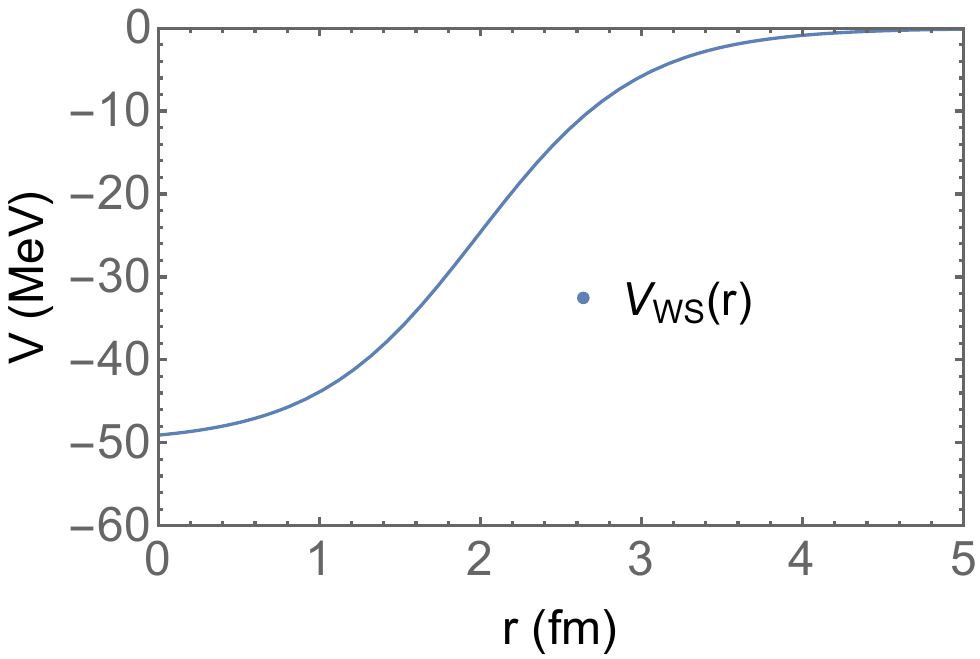}
    \includegraphics[height=5cm,width=0.9\linewidth]{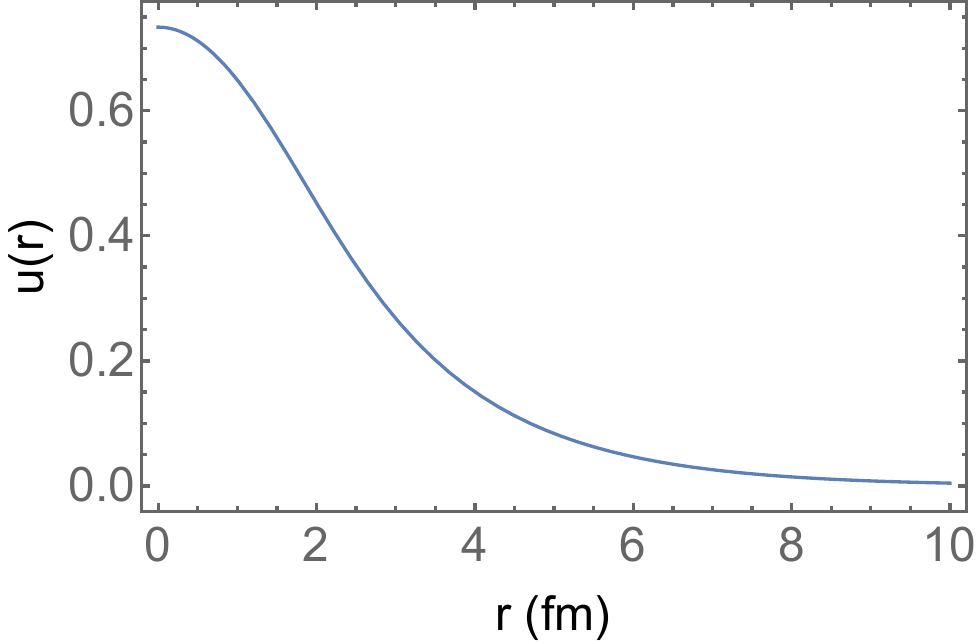}
    \caption{Top: Woods-Saxon potentials $V_{WS}$ for helium-4;
    Bottom: Helium-4 S-wave.}
    \label{fig:WS}
\end{figure}

\subsubsection{Woods-Saxon potential (mean-field approximation)}
For heavier nuclei the use of single particle states in the mean-field approximation is more convenient,
modulo the center of mass motion. Although helium-4 does not qualify as a large nucleus, we will present the analysis for comparison with the K-harmonic method. 
In this case, 
the radial part of Eq.~(\ref{ALPHA1}) will be sought using the independent particle states
\bea
\varphi[r_1, ..., r_4]=\prod_{i=1}^4 \frac{u(r_i)}{r_i}.
\eea
The  reduced $u$ is solution to 
\bea
u^{''}(r)-\frac{m_N}{2\hbar^2}(E_H+V_{WS}(r))u(r)=0,
\eea
in the Woods-Saxon potential
\bea
V_{WS}(r)=-\frac{V_0}{1+e^{(r-R)/a}}\equiv -V_0y(r),
\eea
and  normalized  as
$\int dr \,u^2=1$.
The depth $V_0$, range $R$  and skin $a$ of the potential are fixed to reproduce helium-4 binding energy per particle $\frac 14 E_H=7.1\,{\rm MeV}$ and radius $r_H=1.7\,{\rm fm}$. 
In general, the solution to Eq.~(\ref{ALPHA1}) can be obtained in closed form, in terms of a generalized hypergeometric function~\cite{Flgge1976PracticalQM}
\bea
\label{ALPHA1X}
u(r)=&&Cy(r)^\nu(1-y(r))^\mu\,\nonumber\\
&&\times
{}_2F_1(\mu+\nu, \mu+\nu+1,2\nu+1,y(r)),\nonumber\\
\eea
with $C$ fixed by the normalization.
Here we have set $\mu=i(\gamma^2-\nu^2)^{\frac 13}$ and
$\nu>0$ with
$$\nu^2=\frac{a^2E_Hm_N}{2\hbar^2 },\qquad
\gamma^2=\frac{a^2V_0m_N}{2\hbar^2 }.$$
For light nuclei in general,  $V_0=50\,{\rm MeV}$, $a=0.51\,{\rm fm}$ and $R=r_0A^{\frac 13}$ with $r_0=1.25\,{\rm fm}$. 

In Fig.~\ref{fig:WS} we show the potential for $A=4$ (top), and  
the single particle state  wavefunction for helium-4 (bottom). The numerical binding energy per particle is $\frac 14 E_H=7.1\,{\rm MeV}$, with a  radius of $1.6\,{\rm fm}$  in agreement with the measured charge radius in~\cite{Kalinowski:2018rmf}.

\subsection{Helium-4 EMT}
The way we have presented the derivation  of the deuteron EMT  results in the impulse approximation, can be applied verbatim to helium-4 using the wavefunction (\ref{ALPHA1}), with much simplifications and minor changes, thanks to the absence of a D-wave admixture in helium-4. With this in mind, the results for helium-4 follow from Eq.~(\ref{DR1X}) by inspection,
\begin{widetext}
\bea
\label{DR1XX}
T_H^{00}(k)=&&4T_M(k)\,\bar C_E(k)=\left(m_\alpha+\frac{\vec{k}^2}{4m_\alpha}\right)A^H(k)+\frac{\vec{k}^2}{4m_\alpha}D^{H}(k),
\nonumber\\
T_H^{0j}(k)=&&0,
\nonumber\\
T^{jl}_H(k)=&&4C_M(k)\bar C_E(k)\,\frac{k^jk^l-\delta^{jl}\vec{k}^2}{m_N^2}=
D^H(k)\,
\frac{k^lk^j-\delta^{jl}\vec{k}^2}{4m_\alpha}
\eea
\end{widetext}
with
\bea
A^H(k)&=&\frac{4T_M(k)}{m_\alpha}\bar C_E(k)-\frac{\vec{k}^2}{m_\alpha^3}(T_M+64C_M)\bar C_E(k)
\nonumber\\
&\approx& A(k)\bar C_E(k),\nonumber\\
D_0^H(k)&=&16\frac {m_\alpha}{m_N^2}C_M(k)\bar C_E(k)\approx 
64C(k)\bar C_E(k),\nonumber\\
\eea
with the normalization  $\bar C_E(0)=1$.
\\
\\
{\bf 1. K-harmonic:}
\\
For the reduced S-wave solution (\ref{PHI4X}) we have~\cite{Castilho:1974}
\begin{widetext}
\bea
   \label{BCEKX}
\bar C_E(k) &=&
\int dR\,d\Omega_9\bigg(
\frac 14\sum_{i=1}^4\,e^{-ik\cdot (r_i-R_C)}\,\frac{|u(R)|^2}{\Omega_9}\bigg)=
\int dR\,d\Omega_9
e^{i\frac{\sqrt{3}}{2}\vec{k}\vec{\xi_3}}\,\frac{|u(R)|^2}{\Omega_9}
\nonumber\\
&=&105\int dR \frac{|u(R)|^2}{\left(\frac{1}{2} \sqrt{3} k R \right)^3}j_3\left(\frac{1}{2} \sqrt{3} k R\right),
\eea
\end{widetext}
where we made use of the permutation symmetry of the helium-4
wavefunction expressed in Jacobi coordinates, with $\vec{r}_4-\vec{R}=-\frac{\sqrt{3}}{2}\vec{\xi}_3$.

\begin{figure}
\centering
    \includegraphics[height=5cm,width=0.9\linewidth]{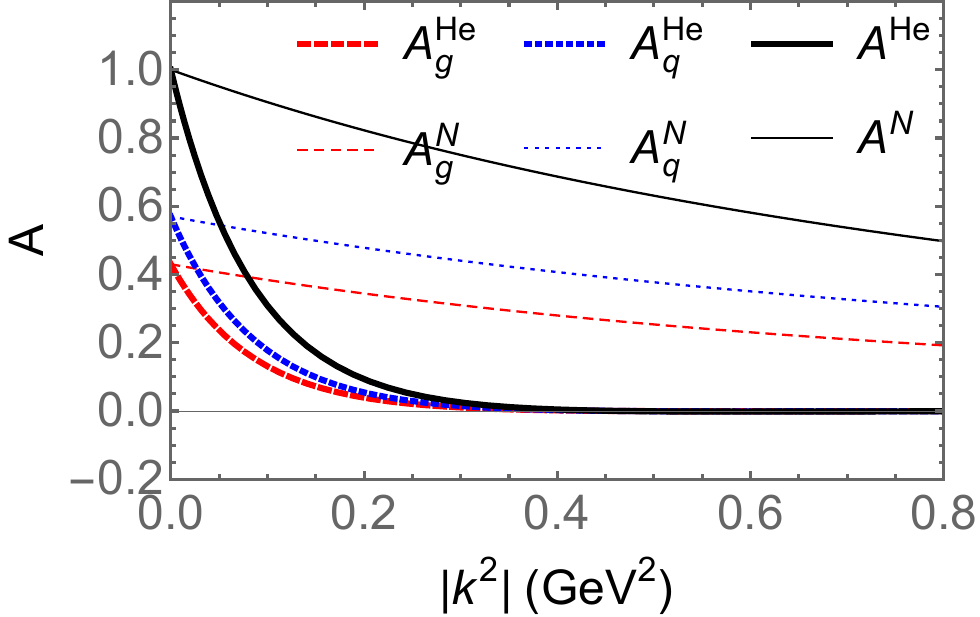}
    \includegraphics[height=5cm,width=0.9\linewidth]{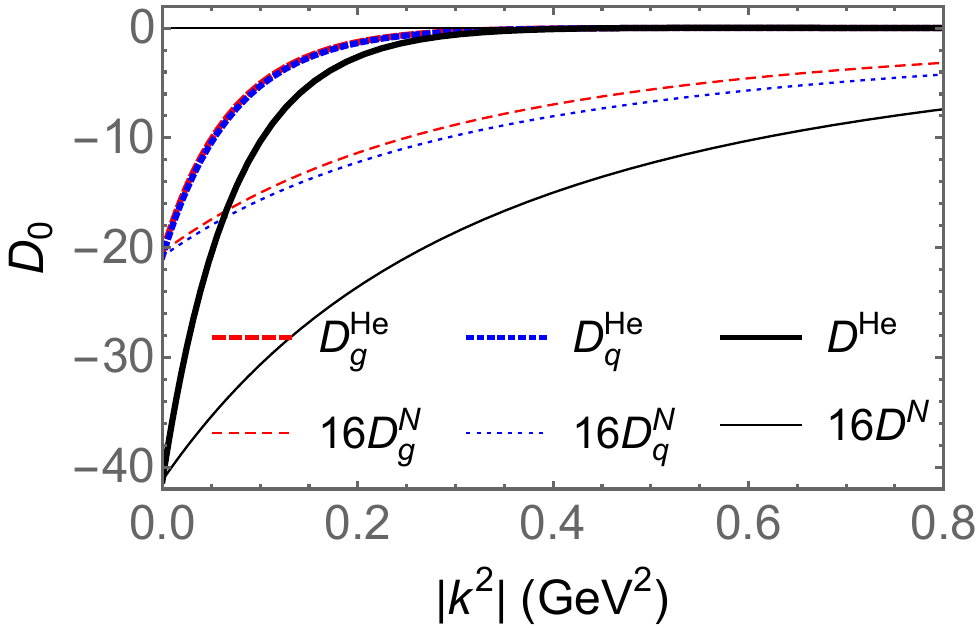}
    \caption{A- and D-form factors for  helium-4 obtained using K-harmonic method with potential $V_3$.}
    \label{fig:AD_K}
\end{figure}

\begin{figure}
\centering
    \includegraphics[height=5cm,width=0.9\linewidth]{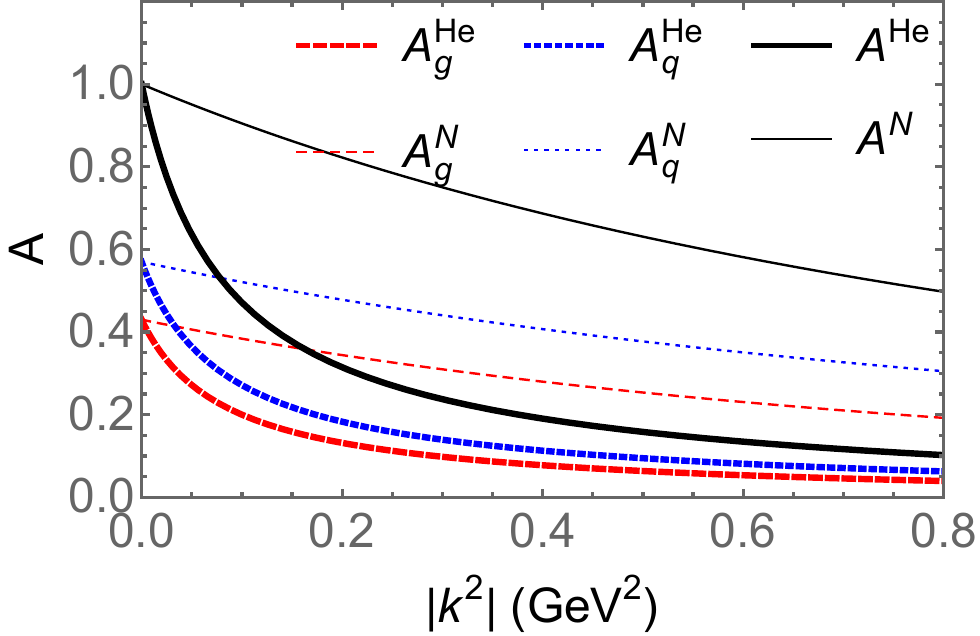}
    \includegraphics[height=5cm,width=0.9\linewidth]{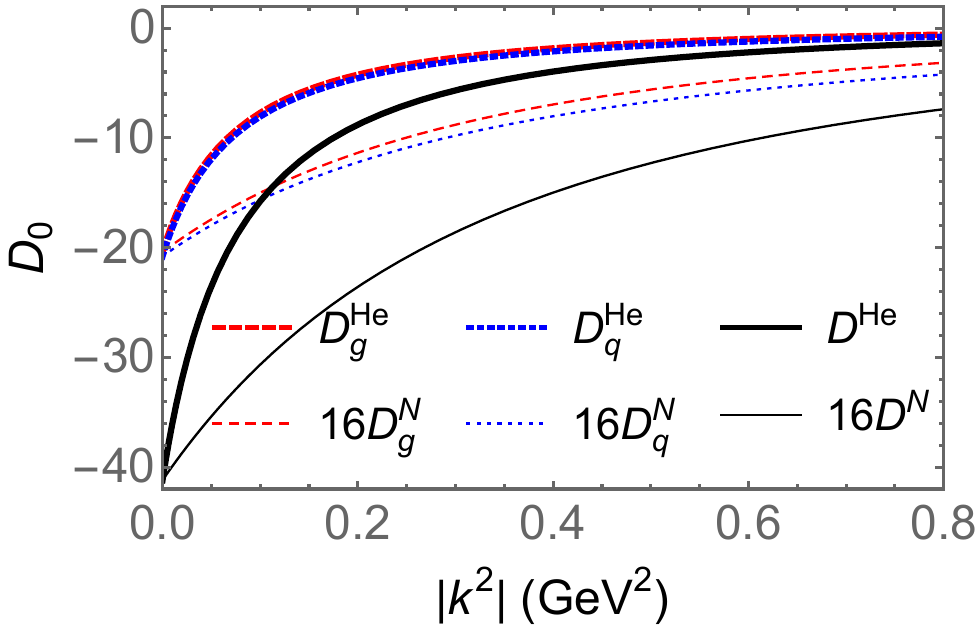}
    \caption{A- and D-form factors for  helium-4 obtained using Woods-Saxon potential.}
    \label{fig:AD_WS}
\end{figure}

{\bf 2. Woods-Saxon potential:}

For the Woods-Saxon potential we have,
\bea
\label{BCEK}
\bar C_E(k)=\int_0^\infty u^2(r)\, j_0({kr}) dr.
\eea
Note that $kr$ instead of $\frac 12 kr$ appears in Eq.~(\ref{BCEK}), as the
the reduced wavefuntions using the Woods-Saxon potential, are coordinated from the center of mass.

\begin{figure}
\centering
    \includegraphics[height=5cm,width=0.9\linewidth]{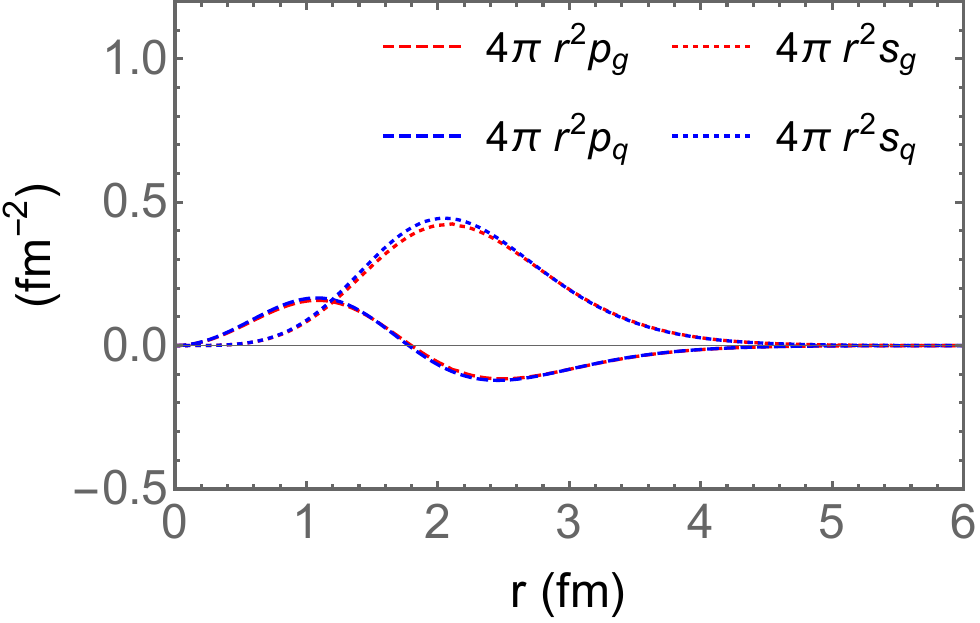}
    \includegraphics[height=5cm,width=0.9\linewidth]{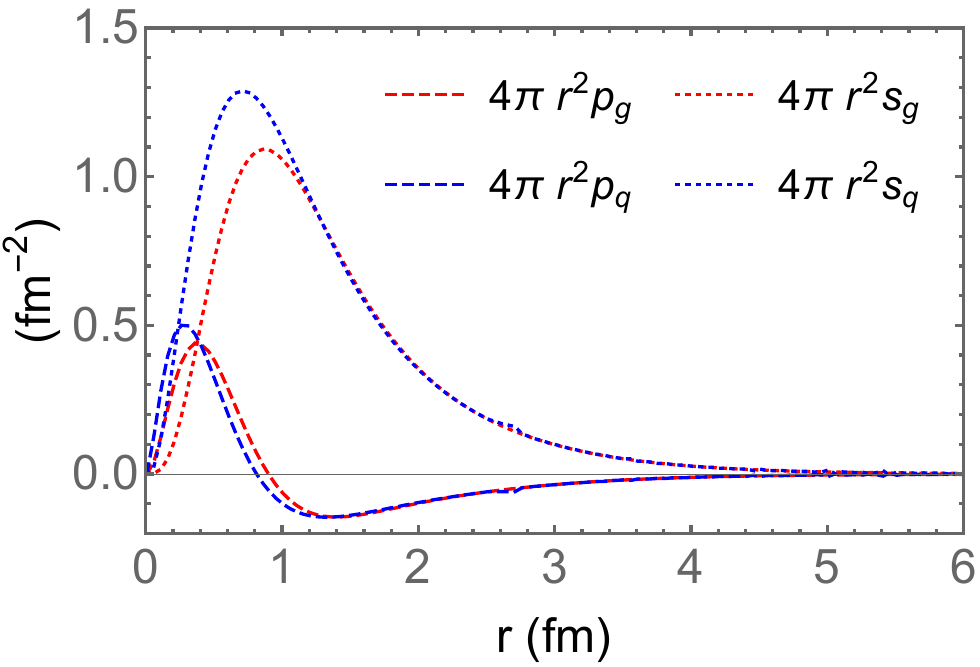}
    \caption{The gluon and quark pressure $p_{g,q}$ and shear $s_{g,q}$ distributions in helium-4 using K-harmonics method(top) and Woods-Saxon potential(bottom), in the impulse approximation.}
    \label{fig:pressure_HE}
\end{figure}

\begin{figure}
\centering
    \includegraphics[height=5cm,width=0.9\linewidth]{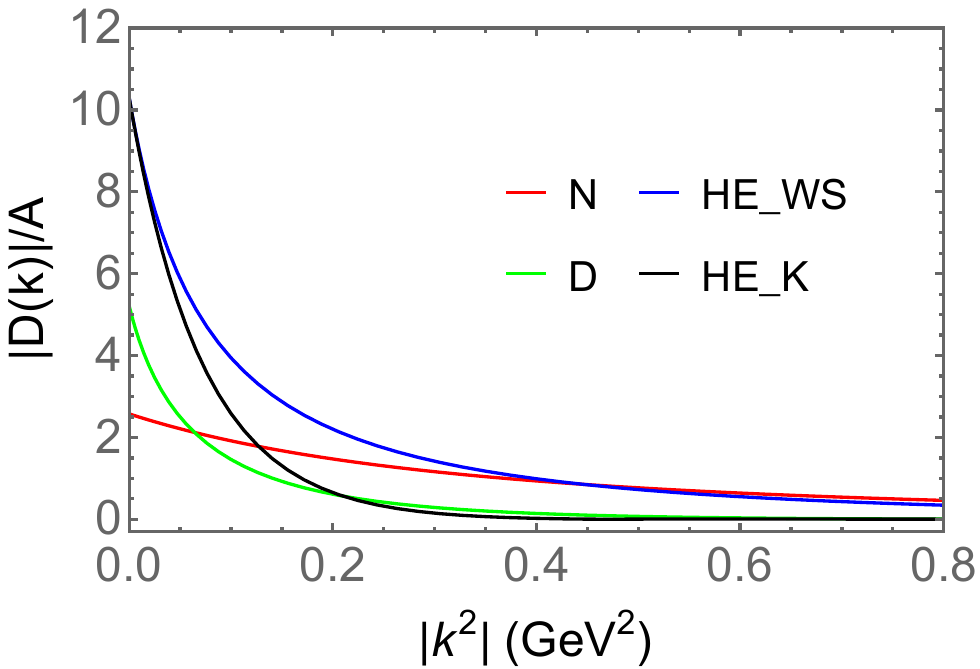} 
  \caption{The spin average D form factor normalized by the baryon number A in the nucleon (N), deuteron (D) and helium-4 (${\rm HE}$) with sub-labels for  K-harmonic with potential $V_3$ and Woods-Saxon.}
    \label{fig:D_normalized}
\end{figure}

In Fig.~\ref{fig:AD_K} we show the A,D form factors for helium-4 using
the K-harmonic method for the wavefunction, with the spurious center of 
mass removed. In Fig.~\ref{fig:AD_WS} we show the A,D form factors for 
helium-4, using the Woods-Saxon potential, without the removal 
of the spurious center of mass. The behavior of the form factors obtained with different approaches
are quantitatively different in the intermediate momentum range, with the
A,D form factors free of the center of mass motion, crossing the zero line at about $k\sim m_N/2$.
The differences between the two constructions 
illustrate the importance of removing the center of mass motion, while describing the form factors for light nuclei. This point is further illustrated in our analysis of the charge form factors in Appendix~\ref{APPA}, where we note the agreement with the potential (\ref{eq:V3}) up to $\frac 23 m_N$, well within the range of validity of our non-relativistic expansion.
The gluon and quark contributions to the pressure $p_{g,q}$ and shear $s_{g,q}$ distributions are shown in Fig.~\ref{fig:pressure_HE}, these results should be reasonable 
for distances $r>2\pi/(\frac 23m_N)\sim 2$ fm.

The results for helium-4, the lightest $0^{++}$ magic nucleus, carry to heavier magic nuclei in the impulse
approximation,  with general $A$ in the Woods-Saxon potential. In particular we have $A^H(0)\approx A^0 A(0)=1$ and
$D^H(0)\approx A^2D(0)$ in the impulse approximation, which is to be compared to the scaling $A^{\frac 73}$ suggested using a liquid drop model~\cite{Polyakov:2002yz}, $A^{2.26}$ using relativistic 
nuclear potentials~\cite{Guzey:2005ba},
and more recently 
$A^{1.7-1.8}$
reported in the generalized Skyrme model~\cite{GarciaMartin-Caro:2023klo}.
In Fig.~\ref{fig:D_normalized} we compare the D-form factor per nucleon,
for the nucleon (red-solid), deuteron (green-solid) and helium-4 with K-harmonic method (black-solid) and Woods-Saxon potential (blue-solid).

\begin{table*}[htbp]
  \centering
  \begin{tabular}{l|c|c|c|c|c|c|c|c|c}
  \toprule
Nuclei 
 & $r^g_S$ & $r^q_S$ & $r^g_M$ & $r^q_M$& $r^g_{S,Q}$ & $r^q_{S,Q}$ & $r^g_{M,Q}$ & $r^q_{M,Q}$ & $r_E$ \\
\hline
proton(experiment) & 1.07~\cite{Duran:2022xag} & -- & 0.76~\cite{Duran:2022xag}& -- & -- & --& --& --&0.84~\cite{ParticleDataGroup:2022pth}\\
\hline
proton(input) &  0.93~\cite{Mamo:2022eui} & 0.82 & 
0.68~\cite{Mamo:2022eui}& 0.60 & -- &-- &-- &-- & 0.8\\
\hline
Deuteron & 2.16 & 2.11 & 2.06 & 2.04 &0.97 & 0.97& 0.97&0.97 & 2.12\,(2.13~\cite{Kalinowski:2018rmf})\\
\hline
Helium-4(K) & 1.80 & 1.76  & 1.69 & 1.67 & --& -- & -- & --& 1.79\,(1.68~\cite{PhysRevC.77.041302})\\
\hline
Helium-4(WS) & 1.79 & 1.75  & 1.68 & 1.66 & --& -- & -- & --& 1.79\,(1.68~\cite{PhysRevC.77.041302})\\
\hline
\hline
  \end{tabular}
\caption{The quark and gluon EMT radii (fm) of light nuclei following from Eq.~(\ref{eq:mass_radius}): scalar radii $r_S$, mass radii $r_M$, $r_{Q,S/M}$  spin averaged quadrupole radii, and $r_E$ charge radii.
For helium-4 we have listed the results from the K-harmonic (K) with potential $V_3$ and Woods-Saxon potential (WS), with the bracketed results referring
to experiment.}
  \label{tab:mass_radius}
\end{table*}

\section{Scalar and mass radii}
\label{SEC6}
We now extend the proton definitions of the scalar radius $r_S$ and the mass radius $r_M$ to light nuclei, by defining the scalar and mass form factors
\bea\label{eq:mass_radius}
\mathbb G_S(k)&=&T^{00}(k)-T^{ii}(k),\nonumber\\
\mathbb G_M(k)&=&T^{00}(k),
\eea
for each of the deuteron and helium-4, with
\bea
r_{S,M}^2=-6\bigg(\frac{d{\rm ln}\,\mathbb G_{S,M}(k)}{d\vec k^2}\bigg)_{\vec k^2=0}.
\eea
The quark and gluon radii in light nuclei are presented in Table.~\ref{tab:mass_radius}, and compared to the charge radii 
following from Appendix~\ref{APPA}
using the same wave-functions.
The quark and gluon separated radii in light nuclei are comparable, owing to the similarity 
of these radii in the nucleon following from Eq.~(\ref{ABFF}). Overall, the difference between the scalar and mass radii seen in the nucleon, persists in light nuclei, with the gluonic scalar radii larger than the mass radii, but
both appear closer to the computed charge radii, in the impulse approximation. A similar observation was made in~\cite{Freese:2022yur} for the light front transverse deuteron size,  in the light cone convolution model.


\section{Conclusions}
\label{SEC7}

We have analyzed the gravitational form factors for the deuteron in  the context of the impulse approximation. The proton and neutron inside the deuteron were assumed non-relativistic, with the 
recoil of the spectator nucleon retained only to linear order. These approximations limit our gravitational form factors to 
momenta of the order of the nucleon mass.

The deuteron gravitational form factors $A^D, Q^D, J^D$ capture the  mass, quadrupole and momentum distributions, supplemented by three additional 
$D^D_0, D^D_2, D^D_3$ form factors reflecting on the standard tensor, spin-tensor and mixed-spin-tensor
contributions, respectively. Using the nucleon gravitational form factors, we have made explicit both the gluonic and fermionic contributions to each of the form factors. This budgeting reflects on the quantum delocalization of the concepts of quarks and gluons, in
constituent bound states at low energy.

The deuteron scalar and mass radii from either the quarks or gluons, are comparable to 
the deuteron electromagnetic radius. In contrast, the spin averaged quadrupole scalar and mass radii carried by the quarks and gluons, are substantially smaller than the deuteron electromagnetic radius.

Our analysis of the deuteron,  readily extends to  helium-4, a much more compact nucleus. 
To describe the $0^{++}$ ground state of helium-4, we have used both the
K-harmonic method where the spurious center of mass motion is explicitly
removed, and a Woods-Saxon potential with the spurious center of mass
motion present. While the radii appears to be similar for both constructions, the
ensuing form factors are substantially different, showing the importance
of removing the spurious center of mass motion.

In the zero momentum limit, the mean-field approximation appears reliable
in the determination of the mass radii, even with the unsubtracted center of mass motion. This observation allows for the extension of  the mean-field approach  to the heavier $0^{++}$
magic nuclei O$^{16}, C^{40}$, .... In particular, Eq.~(\ref{DR1XX}) can be extended to the heavier nuclei case, the spin average D-form factor 
for these heavier nuclei appears to scale as 
$D^A(0)=A\,m_A/m_ND(0)\approx A^2D(0)$ for a large atomic number $A$, in the impulse approximation. Although the non-relativistic reduction holds in heavier nuclei,  fermi motion requires that we include the next-to-leading order corrections in the spectator recoil. Also, exchange current corrections maybe important.

While our analysis has been considerably simplified
by treating the  light nuclei constituents non-relativistically, limiting the range of the invariant form factors to about the nucleon mass, we plan to extend it to the relativistic case at least for the deuteron. Also, our analysis was limited to first order in the recoil of the struck nucleon or core. We plan to pursue the analysis to second order in the spectator recoil momentum, and investigate the importance of the exchange current contributions.

Our construction can be extended to analyze the generalized parton distributions (GPDs) for light nuclei, to understand
the particular role played by the  nucleon pair-interaction, as well as exchange currents. Our gravitational form factors should prove useful for assessing diffractive photo- and electro-production of heavy quarkonia on light nuclei.

The current effort at JLAB to measure near threshold heavy quarkonia production on
nucleons, should be extended to light nuclei, to shed light on how the formation of nuclei may affect our understanding of mass and charge distributions, and the nature of the quantum delocalization of the quarks and gluons in bound states at low energy. Clearly, with the advent of the Electron-Ion Collider (EIC) with higher energy and luminosity, threshold photoproduction of quarkonia such as $J/\Psi, \Upsilon$ on light nuclei, should prove useful for addressing these issues.

\vskip 0.5cm
{\noindent\bf Acknowledgements}

\noindent 
We thank Zein-Eddine Meziani and Bao-Dong Sun for discussions. FH is supported by the National Science Foundation under CAREER Award PHY-1847893.
IZ is supported by the Office of Science, U.S. Department of Energy under Contract  No. DE-FG88ER40388.
This research is also supported in part within the framework of the Quark-Gluon Tomography (QGT) Topical Collaboration, under contract no. DE-SC0023646.

\appendix

\section{Light nuclei charge form factors}
\label{APPA}
To compare the mass radii from the gravitational form factors to the charge radii for light nuclei, we provide here a simple estimate of their charge form factors, also in the impulse approximation. Since data are available, that allows us to gauge the validity of our method.
With this in mind and for a single nucleon, the electromagnetic current  reads
\bea
&&J^\mu_N(k)=\nonumber\\
&&\bar u(p_2)\,e\,
\bigg(F_1(k)\gamma^\mu+F_2(k)\frac{i\sigma^{\mu\nu}q_\nu}{2m_N}\bigg)u(p_1).
\nonumber\\
\eea
$F_1, F_2$ are the Dirac and Pauli form factors,
which are related to the electric and magnetic Sachs form factors as
\bea
G_E(k)&=&F_1(k)-\frac {\vec{k}^2}{4m_N^2}F_2(k),\nonumber\\
G_M(k)&=&F_1(k)+F_2(k).
\eea

The age-old Rosenbluth analysis of the electron scattering data up to $10\,{\rm GeV}^2$ shows that the Sachs form factors for the proton are well approximated by dipoles
\bea
\label{DIPOLE}
G_D(k)&=&\bigg(1+\frac{\vec{k}^2}{0.71}\bigg)^{-2},\nonumber\\
G^p_E(k)&=&G_D(k),\nonumber\\
G_E^n(k)&=&0,\nonumber\\
G_M^{p,n}(k)&=&\mu_{p,n}G_D(k),
\eea
with the $p,n$ empirical magnetic moments $\mu_p=2.79, -1.91$ (in Bohr magnetons).

For completeness, we note that the JLab analysis based on the ratio of the polarization of the  scattered proton, shows $G^p_E$ falling faster than $G^p_M$~\cite{JeffersonLabHallA:2001qqe}  
\bea
G_E^p(k)=(1-0.13\,(\vec k^2-0.04))\,G_D(k).
\eea

\begin{figure}
\centering
    \includegraphics[height=5cm,width=0.9\linewidth]{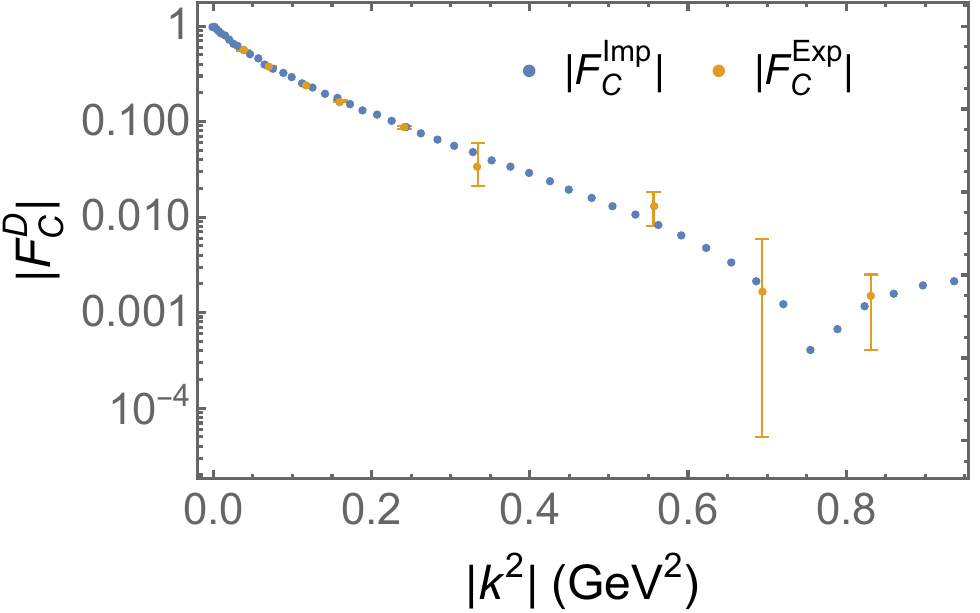}
    \includegraphics[height=5cm,width=0.9\linewidth]{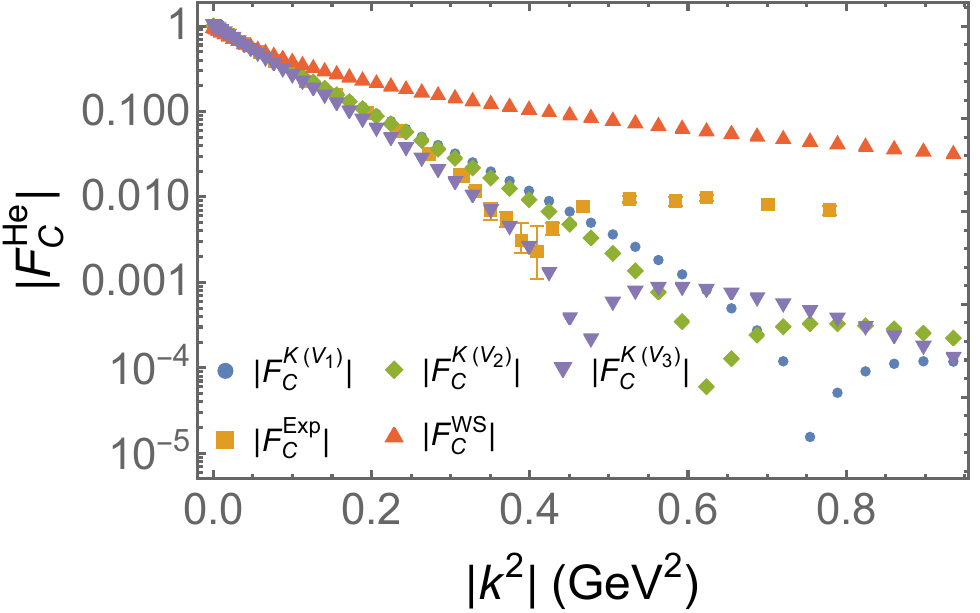}
    \caption{Top: Deuteron electric charge form factors 
    in the impulse approximation compared to experiment~\cite{Garcon:1993vm}.
    Bottom: Helium-4 electric charge form factors using the K-harmonic method 
    with the potentials in Eqs.~(\ref{CAST}-\ref{eq:V3}) and the Woods-Saxon potential, compared to 
    experiment~\cite{Frosch:1967pz}.}
    \label{fig:FCD}
\end{figure}

The leading non-relativistic and recoil contributions to the nucleon
charge form factor, are
\begin{widetext}
\bea
\label{J0N}
eJ^0_{N}(k)=e\,\bigg(G_E(k)
+\frac{(\sigma\times ik)\cdot P}{4m_N^2} (2G_M(k)-G_E(k)) +{\cal O}\bigg(\frac{\vec{k}^4}{m_N^4}, \frac{P^2}{m_N^2}\bigg)\bigg).
\eea
\end{widetext}
We now proceed to use Eq.~(\ref{J0N}) for the modifications to the charge  density in light nuclei, using the impulse approximation.
\\
\\
{\bf Deuteron charge form factor:}
\\
Since the deuteron wavefunction is symmetric under spin exchange and independently space exchange
of $p,n$, it follows that only the singlet combination of electric and magnetic form factors contribute to the deuteron charge form factor (\ref{J0N}) through the substitution 
\bea
\label{SINGLET}
G^S_E(k)&=& \frac{G^p_{E}(k) +G^n_{E} (k)}{2},\nonumber\\
G^S_M(k)&=& \frac{G^p_{M}(k) +G^n_{M} (k)}{2},
\eea
with $\sigma\rightarrow S$.
With this in mind, and
using some of the matrix elements developed for the EMT form factors earlier, we obtain for the charge density in the impulse approximation
\begin{widetext}
\bea
\label{J0X}
J^0_{D}(k, m', m)=&&
2G^S_E(k)\,(C_E(k)\delta_{m'm}-2C_Q(k)\langle m'|(S\cdot \hat k)^2-\frac 13 S^2|m\rangle)\nonumber\\
-&&\frac{\vec{k}^2}{2m_N^2}\,(2G^S_M(k)-G^S_E(k))\,
(D^{SP}_0(k)\delta_{m'm}+(D^{SP}_2(k)+2D^{SP}_3(k))
\langle m'|Q^{ij}\hat k^i\hat k^j|m\rangle)
\nonumber\\
&=&F^D_C(k)\delta_{m'm}+F^D_Q(k)\frac{k^\alpha k^\beta}{2m_D^2}\langle m'|Q^{\alpha\beta}|m\rangle.
\eea
\end{widetext}
The squared electric charge radius of the deuteron is the sum of the nucleon,  plus the intrinsic $C_E$-form factor contribution
\bea
\langle r^2\rangle_D=\langle r^2\rangle_N +
\langle r^2\rangle_{C_E}.
\eea
The results for the electric charge form factor for the deuteron $|F_C^D|$  in Eq.~(\ref{J0X})  are shown in Fig.~\ref{fig:FCD} (top), and compared to the empirical data in~\cite{Bekzhanov:2013yoa}. 
The impulse approximation works reasonably well in this momentum range, although the diffractive dip is slightly off to the right of the empirical values.
The diffractive pattern with a first zero at about $k^2_D\approx 0.75\,{\rm GeV}^2$, reflects on the good deuteron S-wavefunction from the soft Reid potential in Fig.~\ref{fig:REID}, with a peak at $a_D\approx 1.5\,{\rm fm}$ (diffraction disc size). 
\\
\\
{\bf Helium-4 charge form factor:}
\\
For helium-4, the charge density in the impulse approximation reads
\bea
\label{J0Z}
J^0_{He}(k)=4G_E^S(k)\,
\bar C_E(k)= 2F_C^{\text{He}}(k),
\eea
where the analogue of the singlet  substitution (\ref{SINGLET}) applies,
because of the spin and space symmetry of the underlying wavefunction. Note that at  low momentum transfer (\ref{J0Z}) simplifies
\bea
F_C^{\text{He}}(k)\approx F^D_C(k).\nonumber\\
\eea
The squared electric charge radius of 
helium-4 is the sum of the nucleon plus the
intrinsic $\bar{\rm C}_E$ form factor contribution
\bea
\langle r^2\rangle_H=\langle r^2\rangle_N +
\langle r^2\rangle_{{\bar C}_E}
\eea
The results for the electric charge form factor for helium-4 $|F_C^{He}|$  in Eq.~(\ref{J0Z}),  are shown in Fig.~\ref{fig:FCD} (bottom) for the Woods-Saxon potential (red-triangle) and the K-harmonic with potential $V_1$ (dotted-blue), $V_2$ (green-diamond) $V_3$ (purple-triangle), and compared to the empirical data (yellow-square) in~\cite{Frosch:1967pz}. Note that the diffractive minima following 
from $V_1-V_3$ are different.
The  corresponding wavefunctions shown in Fig.~\ref{fig:WS4} are
similar but not identical, especially  in the small $R$ region, which is at the origin
of the difference in the $k>m_N/2$ region.
The result with Woods-Saxon potential does not display a diffraction pattern, since the single-particle wavefunction in Fig.~\ref{fig:WS} shows no  plateau or disc. The K-harmonic does, and the discrepancy between the measured diffractive minimum and the K-harmonic minimum, could be narrowed by optimizing the potential choice in Eq.~(\ref{CAST}). Away from the diffractive minima, the K-harmonic electric form factor and the measured one are comparable in magnitude.

\section{Pressure and shear force}
\label{app:pre}
Following~\cite{Polyakov:2019lbq}, the stress tensor defined by the ij-components of the EMT is defined as
($a=q,g$)
\begin{widetext}    
\bea \label{eq:static_EMT}
T^{jl}_a(\vec r, m^\prime,m) &=& \int {d^3 k \over (2\pi)^3 }\frac{m_D}{E} e^{-i k \cdot r}
\langle +\frac{k}{2} m^\prime \, |{\hat T}^{jl}_a(0)|-\frac{k}{2} m\rangle \ \nonumber\\
&=&
(p_0(r) \delta^{jl}+s_0(r)Y_2^{jl})\delta_{m'm}
+ p_2(r)\langle m'|Q^{ij}|m\rangle
+
2s_2(r) \langle m'|\hat{Q}^{lp}Y_2^{pj}+
\hat Q^{jp}Y_{2}^{pl} -\delta^{jl} \hat Q^{pq}Y_{2}^{pq}|m\rangle 
\nonumber \\
&+& 
\langle m'|Q^{\alpha\beta}|m\rangle 
\hat{\partial}_\alpha\hat{\partial}_\beta[p_3(r)\delta^{jl}+s_3(r)Y_2^{jl}],
\eea
\end{widetext}
with $Y_2^{jl}=\hat{r}^l\hat{r}^j-\frac{1}{3}\delta^{jl}$. The pressure and shear force follow as
\bea
p_{i}&=&\frac{1}{3}\frac{1}{r^2}\frac{d}{dr}r^2\frac{d}{dr}\tilde{D}_{i}(r),\nonumber \\
s_{i}&=&-\frac{1}{2}r\frac{d}{dr}\frac{1}{r}\frac{d}{dr}\tilde{D}_{i}(r),
\eea
Here $\tilde{D}_{i}$ are the Fourier transform of the deuteron form factors $D_{i}$ defined in Eq.~(\ref{DR1X}),
\bea
\tilde{D}_{0,2,3}(r)=\int {d^3 k \over 2E(2\pi)^3 } e^{-i k \cdot  r}D_{0,2,3}(k), 
\eea
For helium-4 without the small D-wave admixture, the pressure and shear force receive contribution only from  $D_0$.

\bibliographystyle{apsrev4-1}
\bibliography{refdeuteron}
\end{document}